\documentclass[aps, prl, reprint, superscriptaddress, floatfix]{revtex4-2}
\usepackage{bm}
\usepackage{graphicx}
\usepackage{xfrac}
\usepackage{float}
\usepackage{mathtools}
\usepackage[dvipsnames]{xcolor}
\definecolor{prlblue}{HTML}{211e7e}

\definecolor{darkgreen}{HTML}{000000}
\definecolor{darkgreen2}{HTML}{000000}

\usepackage[hidelinks,breaklinks,colorlinks = true,urlcolor = prlblue,citecolor = prlblue, linkcolor = black,]{hyperref}
\usepackage{amsthm,gensymb}
\usepackage{upgreek}
\usepackage{mathptmx}
\usepackage{chemformula}
\usepackage{derivative}
\usepackage[version=4]{mhchem}

\begin{document}

%Affiliations
\newcommand*{\cpfs}{Max Planck Institute for Chemical Physics of Solids, 01187 Dresden, Germany}

\newcommand*{\lsi}{Laboratoire des Solides Irradi\'{e}s, CEA/DRF/IRAMIS, \'{E}cole Polytechnique, CNRS, Institut Polytechnique de Paris, 91128 Palaiseau, France}

\newcommand*{\standrews}{School of Physics and Astronomy, University of St Andrews, St Andrews KY16 9SS, United Kingdom}

\newcommand*{\stanfordslac}{Stanford Institute for Materials and Energy Sciences,
SLAC National Accelerator Laboratory, Menlo Park, CA 94025, USA}

\newcommand*{\stanfordphysics}{Department of Physics, Stanford University, Stanford, CA 94305, USA}

\newcommand*{\stanfordmaterials}{Department of Materials Sciences, Stanford University, Stanford, CA 94305, USA}

\newcommand*{\stanfordapplphysics}{Department of Applied Physics, Stanford University, Stanford, CA 94305, USA}

% to easily write the dose
\newcommand{\dose}[1]{#1~C/cm\textsuperscript{2}}
\newcommand{\Tc}{$T_\mathrm{c}$}

\title{Disorder-induced suppression of superconductivity in infinite-layer nickelates }

% Disorder-dependent suppression of superconductivity in infinite-layer nickelates 

% \title{Controllably induced disorder reveals unconventional pairing symmetry in superconducting infinite-layer nickelates} 

% Author List
\author{Abhishek Ranna}
% \email[]{abhishek.ranna@cpfs.mpg.de}
\affiliation{\cpfs}

\author{Romain Grasset}
\affiliation{\lsi}

\author{Martin Gonzalez}
\affiliation{\stanfordslac}
\affiliation{\stanfordmaterials}

\author{Kyuho Lee}
\affiliation{\stanfordslac}
\affiliation{\stanfordphysics}

\author{Bai Yang Wang}
\affiliation{\stanfordslac}
\affiliation{\stanfordphysics}

\author{Edgar Abarca Morales}
\affiliation{\cpfs}

\author{Florian Theuss}
\affiliation{\stanfordslac}
\affiliation{\stanfordapplphysics}

\author{Zuzanna H. Filipiak}
\affiliation{\cpfs}
\affiliation{\standrews}

\author{Michal Moravec}
\affiliation{\cpfs}
\affiliation{\standrews}

\author{Marcin Konczykowski}
\affiliation{\lsi}

\author{Harold Y. Hwang}
\affiliation{\stanfordslac}
\affiliation{\stanfordapplphysics}

\author{Andrew P. Mackenzie}
\affiliation{\cpfs}
\affiliation{\standrews}

\author{Berit H. Goodge}
\email[]{berit.goodge@cpfs.mpg.de}
\affiliation{\cpfs}

%\date{\today}

\begin{abstract}

The pairing symmetry of superconducting infinite-layer nickelates is a fundamental yet experimentally challenging question. We employ high-energy electron irradiation to induce disorder in superconducting Nd$_{0.825}$Sr$_{0.175}$NiO$_{2}$ thin films and examine the impact of pair-breaking defects on superconductivity and elucidate the nature of the superconducting gap.
Our measurements reveal a complete suppression of superconductivity with increasing disorder, suggesting an unconventional, sign-changing order parameter.
\end{abstract}

\maketitle

% \section{Introduction}

Superconductivity in hole-doped infinite-layer rare-earth nickelates was realized two decades after their theoretical prediction \cite{anisimov_1999_prb} through a combination of crystalline thin-film synthesis with topotactic chemical reduction to stabilise the requisite nickel 3$d^9$ valence \cite{li_superconductivity_2019}, providing an experimental platform to explore a close electronic and structural analogue to high-\textit{T}\textsubscript{c} cuprates \cite{Bednorz1986_cuprate, WangLeeGoodge-Review}. 
Despite the similarities between the phase diagrams of the Ni and Cu-based systems, several key distinctions between their electronic landscapes and hybridization have been identified \cite{Lee2004_Ni_is_not_Cu, Hepting2020_parent_electronic_structure, goodge2021doping}.

One fundamental and outstanding question is the symmetry of the superconducting order parameter in infinite-layer nickelates.
Experimentally, this has remained challenging to address, hindered largely by the thin film geometry and sample synthesis. 
An early single particle tunneling experiment, for example, found both $s$-wave and $d$-wave-like gap structures across a single film, suggesting limitations from sample surface quality \cite{gu2020single}.
Temperature-dependent measurements of superfluid density by mutual inductance \cite{Harvey2022MutualInductance} and tunnel-diode-oscillator \cite{chow2023pairing} approaches have so far led to differing interpretations, particularly complicated by the low-temperature behaviour in the Nd-based nickelates. 
Recent optical measurements by terahertz spectroscopy exhibit responses which could be attributed to $d$-wave pairing \cite{cheng2024evidence}.
Other conventional experimental probes of pairing symmetry such as thermal conductivity and specific heat are generally not compatible with a very thin ($<10$ nm) film geometry \cite{Grissonnanche2024_seebeck}. 

Alternatively, the dependence of superconductivity on local \textcolor{darkgreen}{non-magnetic} pair-breaking disorder can be an effective way to distinguish between nodal and fully gapped systems following the formulation of Abrikosov and Gor'kov \cite{abrikosov_gorkov, Radtke1993, Mackenzie2003_rmp_review}. 
In some materials, non-magnetic chemical impurities can be introduced during synthesis as a way to tune disorder \cite{Alloul_YBCO_Zn_1991,Fukuzumi1996,Karpinska_LSCO_Zn_2000, Mackenzie_SRO_1998}.
Given the myriad synthetic challenges already posed by the infinite-layer nickelates, however, a more practical approach in this system is post-synthetic introduction of pair-breaking disorder into the highest quality superconducting films.

Here, we harness high-energy electron irradiation to controllably induce point-like disorder in optimally hole-doped superconducting Nd$_{0.825}$Sr$_{0.175}$NiO$_{2}$ (NSNO) thin films. 
Primary electrons with megavolt-range energies undergo nearly elastic collisions with atoms in the sample, displacing single atoms from their crystalline positions to create vacancy-interstitial pairs \cite{Giapintzakis1992_YCBO_irradiation, konczykowskiYBCO_1991}. 
Importantly, however, the transferred energy is not sufficient for the displaced atom to create additional defects (cascade processes), such that the resulting disorder is considered to be comprised of single point-like rather than extended defects \cite{Alessi2023_Electron_irradiation, sunko2020controlled} (for further discussion see Supplemental Note A). 
With this ability to systematically tune point-like disorder, high-energy electron irradiation has been used as an effective platform to investigate the pairing symmetries of a wide range of superconducting compounds, such as cuprates \cite{konczykowskiYBCO_1991, legris1993, Rullier2003, Openov2005}, ruthenates \cite{ruf_noad2024_SRO}, organic superconductors \cite{analytis2006effect}, iron pnictides \cite{prozorov2014effect},  kagome systems \cite{Roppongi2023}, doped SrTiO$_3$, \cite{Xiao2015_dopedSTO} \textcolor{darkgreen}{and topological superconductors \cite{Ghimire2024_LaNiGa2}.}

Our measurements show that increasing disorder systematically and reproducibly suppresses superconductivity in infinite-layer nickelates, pointing to a sign-changing superconducting order parameter.
Electrical resistivity also increases with disorder in all samples, but we reveal quantitative variation in the evolution of parameters such as residual resistivity and the normal state linear-in-temperature slope, which we attribute to additional sources of scattering which cannot be fully captured by simplified considerations.
These results point to unconventional nickelate superconductivity and also provide a uniquely systematic study of disorder signatures in these materials, instructive for both physical insights as well as ongoing materials optimization efforts.

Thin films of infinite-layer Nd$_{0.825}$Sr$_{0.175}$NiO$_{2}$ were obtained by first synthesizing perovskite precursor Nd$_{0.825}$Sr$_{0.175}$NiO$_{3}$ thin films on (LaAlO$_3$)$_{0.3}$(Sr$_2$TaAlO$_6$)$_{0.7}$ (LSAT) (001)-oriented substrates (lattice constant = 3.868 {\AA}) by pulsed-laser deposition using a KrF eximer laser ($\lambda = 248$ nm) followed by a topotactic reduction using \ce{CaH2}.
All films used here are $\sim$5 nm thick and capped with a few unit cells of epitaxial \ce{SrTiO3} which act as a supporting and protective layer during the apical oxygen removal process. 
Further details of the synthesis are discussed in \cite{li_superconductivity_2019, Lee2020_APL, Lee2023_normal_state, Gonzalez2024}.
Prior to electron irradiation, sample substrates were back-polished with diamond grit lapping paper to reduce the overall thickness to $<$300 {\textmu}m, which reduces heating effects and attenuation of the high energy electron beam during irradiation. 

\begin{figure}[t]
    \centering
    \includegraphics[width = \columnwidth]{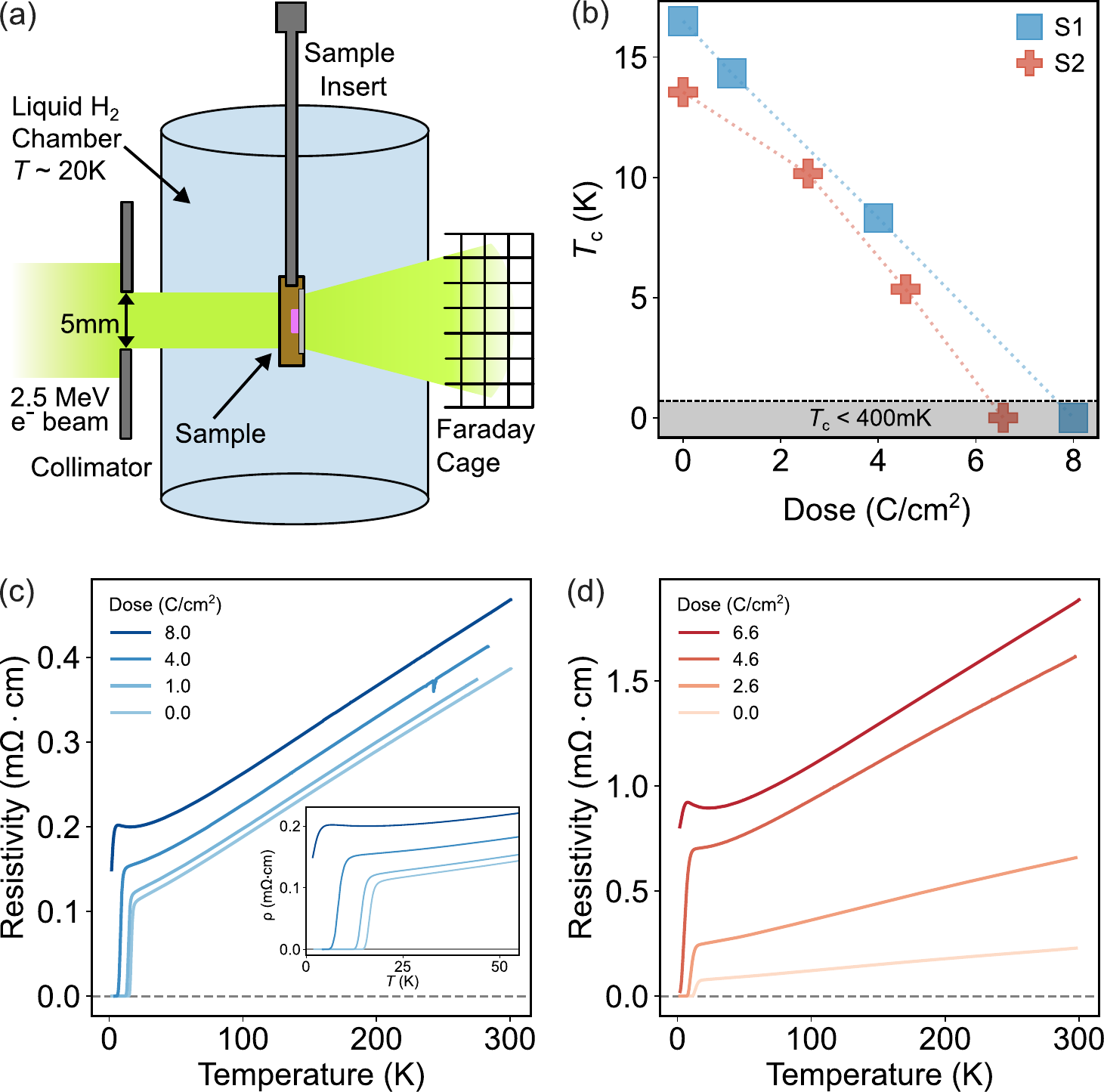}
    \caption{
    (a) Schematic of the high energy electron irradiation experiment: the sample (purple) is mounted on an insert, placed in a flowing liquid hydrogen chamber (blue), and rastered with a 2.5 MeV electron beam (green). A Faraday cage measures the transmitted current to monitor the accumulated dose. (b) Evolution of {\Tc} with accumulated 2.5 MeV electron irradiation dose for multiple NSNO thin films. (c) Resistivity as a function of temperature of NSNO thin film S1 with electron irradiation at 2.5 MeV with progressive total doses of 0 (pristine), 1.0, 4.0 and \dose{8.0}. Inset: The superconducting transition is suppressed with increasing dose. (d) Resistivity as a function of temperature of a second NSNO thin film S2 for 2.5 MeV electron irradiation with progressive total doses of 0 (pristine), 2.6, 4.6 and \dose{6.6}.
    }
    \label{fig1}
\end{figure}

% \par
% \subsection*{Electron irradiation} 
Electron irradiation was performed using 2.5 MeV electrons at the SIRIUS Pelletron accelerator at Laboratoire des Solides Irradi\'{e}s, \'{E}cole Polytechnique  (Palaiseau, France). 
Electrons with this energy have a large penetration depth on the order of millimeters, ensuring a homogeneous and complete transmission of the beam through the entire thin film sample \cite{Alessi2023_Electron_irradiation}.
During irradiation, the samples are held at $\sim$22 K by a flow of liquid hydrogen within the CRYO-1 irradiation cell to immobilise irradiation-induced defects upon formation and remove the heat produced by relativistic electrons upon collision with the atoms (Fig. \ref{fig1}a). 
The irradiation dose is monitored by measuring the current from transmitted electrons with a Faraday cage placed behind the sample stage.

Total irradiation doses were accumulated over multiple runs, between which samples were periodically removed to perform electronic transport measurements.
It is expected that some fraction of the defects induced by irradiation are lost (e.g., recombined through annealing) when the sample is warmed to room temperature for external measurements \cite{Alessi2023_Electron_irradiation,sunko2020controlled,  ruf_noad2024_SRO}.  
In our study, this annealing makes quantitative correlation between defect density and electron dose challenging, but the relative trends between irradiation induced disorder and various properties measured by electronic transport are robust. 
\textcolor{darkgreen}{We confirm by x-ray diffraction measurements and atomic-resolution scanning transmission electron microscopy that the overall structure and crystallinity of the infinite-layer phase are preserved following electron irradiation, finding no evidence of additional structural changes (Supplemental Notes D and E).}

\begin{figure*}
    \centering
    \includegraphics[width = 0.77\textwidth]{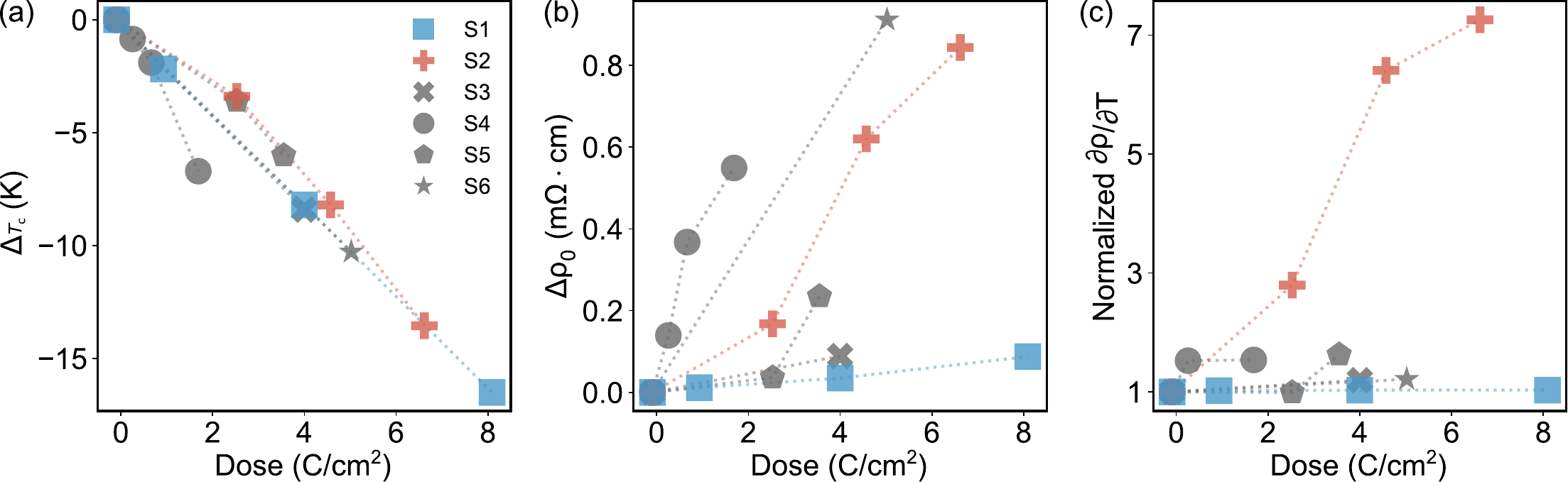}
    \caption{Evolution of transport properties for several NSNO thin films with increasing dose, including (a) change in superconducting transition {\Tc}, (b) change in residual resistivity with respect to the pristine values ($\Delta \rho_0$ as defined in the main text), and (c) normalized \textit{T}-linear normal state slope $\pdv{\rho}/{T}$.}
    \label{fig2}
\end{figure*}

% \subsection*{Superconductivity} 
Figure \ref{fig1}b-d shows the effect of accumulated electron irradiation on two superconducting NSNO thin films S1 and S2, namely the decrease of superconducting transition temperature  ({\Tc}) and increase in resistivity ($\rho$) with increasing dose. 
Progressive measurements of temperature-dependent electronic transport of S1 and S2 are shown in Figs. \ref{fig1}c and d, both illustrating a characteristic evolution with decreasing {\Tc} from the as-synthesized or pristine film (0 C/cm$^2$) across total accumulated doses up to 8 C/cm$^2$. 
Here we define {\Tc} as the temperature where the resistivity falls to 50\% within the downturn; details of extracting {\Tc} and other parameters are discussed in Supplemental Fig. S2.
% The inset shows a clear decrease in {\Tc} with increasing dose. 
The sharpness of the superconducting transition shows negligible changes between measurements, confirming both the point-like nature of the induced defects and the homogeneity of their distribution across an entire thin film sample (Supplemental Fig. S3) \cite{Alessi2023_Electron_irradiation}.
Similar {\Tc} suppression is observed across several irradiated samples (Supplemental Figs. S4, S5). 

As the disorder increases with additional irradiation, low-temperature resistive upturns appear for both S1 and S2. 
Similar upturns were observed in electron irradiation studies of YBa$_2$Cu$_3$O$_{7-\delta}$ (YBCO), where they have been discussed in the context of combined single impurity scattering and localization effects \cite{Rullier2001, Rullier2003}.
Low-temperature resistive upturns are also observed in as-synthesized under- and over-doped NSNO films of similar quality \cite{Lee2023_normal_state}.
We note, however, that the resistive upturn we observe here is dependent on increasing disorder (similar to the pristine overdoped and distinct from pristine underdoped NSNO films) but not affected by applied magnetic fields (Supplemental Fig. S6, similar to pristine underdoped and distinct from pristine overdoped films).
Further investigations with more densely-sampled irradiation doses may elucidate the microscopic origin of the upturn in the specific case here.

Full suppression of the superconducting state (i.e., {\Tc} $<$ 400 mK) is achieved in both samples with the total accumulated doses of 8.0 and 6.6 C/cm$^2$ for S1 and S2, respectively (Fig. \ref{fig1}b). 
\textcolor{darkgreen}{
The systematic and consistent suppression of {\Tc} at a rate of $\sim$2 K per C/cm$^2$ irradiation across all samples (Fig. \ref{fig2}a) demonstrates this method as a reliable and controllable way of tuning point-like disorder in these films.
}

Also accompanying the decrease in superconducting transition temperature is a systematic increase in resistivity with accumulated dose, albeit at notably different rates between films (Supplemental Fig. S5b).
The residual resistivity (resistivity just before the superconducting downturn) $\rho_0$ is commonly used as a convenient parameter to quantify the degree of intrinsic disorder in a system \cite{sun1994electron,Fukuzumi1996, Mackenzie_SRO_1998}.
In many real materials, however, other extrinsic factors such as extended defects or epitaxial strain can also contribute to additional scattering. 
Here, we define the residual resistivity of pristine (pre-irradiated) films $\rho_0^{\mathrm{pristine}}$ as a baseline for all the existing disorder and/or defects present in as-synthesized samples, and use the change in residual resistivity at a given dose $\Delta \rho_0 = \rho_0 - \rho_0^{\mathrm{pristine}}$ as a proxy for the controllably induced disorder \cite{Rullier2003, prozorov2014effect, sunko2020controlled, ruf_noad2024_SRO}.
Unlike $\Delta${\Tc}, $\Delta \rho_0$ per dose is highly sample dependent (Fig. \ref{fig2}b), \textcolor{darkgreen}{suggesting additional scattering contributions, for example within extended defects, which are also affected by irradiation-induced disorder.}

\textcolor{darkgreen}{This is further hinted by the evolution of the normal state slope with increasing disorder.}
The linear-in-temperature resistivity in optimally doped nickelates is phenomenologically similar to superconducting cuprates and other materials also dubbed `strange metals' \cite{Lee2023_normal_state, Bruin_Mackenzie2013_Science, Hartnoll_Mackenzie2022_RMP, Varma2020_T-Linear_RMP, martin1990normal, doiron2009correlation, Lohneysen1994}.
Strange metallic resistivity has a general form $\rho(T) \sim \rho(T=0) +AT$ \cite{Anderson1988_VBS_Transport}, where the slope $A$ = $\pdv{\rho}/{T}$ captures (in principle) the scaling $\rho \propto k_B T$ within the frameworks of linear scattering rates \cite{Orenstein1990, Marel2003, Zaanen2004} and the Drude approximation. 
% This $T$-linear resistivity points to scattering rates $\frac{1}{\tau_{tr}} \propto k_B T/\hbar$ .
% In the paradigm of the simplistic Drude approximation $\sigma \propto \tau_{tr}$ hence $\rho$ scales with $k_B T /hbar$. 
% The normal state slope, $\pdv{\rho}/{T}$ captures this scaling. 
% However, any quantitative analysis involving $A$ requires accurate information and mapping of the Fermi surface \cite{Hartnoll_Mackenzie2022_RMP}. 
It was previously shown that optimally doped nickelates exhibit an intriguingly similar normal state slope to hole-doped La$_{2-x}$Sr$_x$CuO$_4$ (LSCO) \cite{Takagi1992_LSCO, Boebinger1996_LSCO, Cooper2009_LSCO, Lee2023_normal_state}.
%Reviewer 2 points here 
Here, we find the normal state slope $\pdv{\rho}/{T}$ in sample S1 remains consistent with increasing disorder, reminiscent of substitutional disorder studies in LSCO \cite{Fukuzumi1996}, both of which follow Matthiessen's rule that the total scattering rate is a sum of all forms of scattering assuming all scattering processes are independent. 
\textcolor{darkgreen2}{Similar evolution has also been observed upon electron irradiation in iron pnictide superconductors, ascribed to increased impurity (elastic) scattering \cite{Mizukami2014_iron_pnictide}.}
Notably, however, in sample S2 we observe a substantial change in $\pdv{\rho}/{T}$ with increasing electron dose,
in an apparent violation of Matthiessen's rule.

\textcolor{darkgreen}{Broadly, we find that samples which exhibit larger changes in normal state slope $\pdv{\rho}/{T}$ (i.e., violate Matthiessen's rule) also tend to exhibit larger changes in $\rho_0$ and $\Delta \rho_0$ with increasing disorder  (Fig. \ref{fig2}b,c; Supplemental Fig. S5),
suggesting that these effects may be linked.
We therefore posit that this apparent violation of Matthiessen's rule may be related to the presence of other sources of scattering}, such as sparse extended defects or other types of native disorder in the as-synthesized films (e.g., solid solution cation disorder, surface or interface scattering in the thin film geometry), which cannot be fully accounted for by our measurements at this stage.
We emphasize, however, that all films presented in this study are among the highest quality superconducting nickelates produced to date, as evidenced by their relatively low pristine resistivity and $T$-linear slopes $\pdv{\rho}/{T}$ down to lowest temperatures prior to irradiation \cite{Lee2023_normal_state} and confirmed by atomic-scale structural characterization (Supplemental Fig. S7).
% These results suggest that the highest quality nickelate thin films (e.g., S1) are ... Regarding the inconsistency with Matthiessen's rule sometimes, I think this is basically saying that while our films are getting better, they are still at the verge of being in the limit of simple perturbative disorder as grown. Whether this is intrinsic, or from remaining extended defects, is unclear to me.
These results provide clear caution against the over-interpretation of macroscopic electronic transport quantities, e.g. $\rho_0$, in the context of microscopic physics in these -- and likely other -- material systems.

\textcolor{darkgreen}{Samples S1 and S2 thus represent the spread of disorder-dependent evolution observed in NSNO films studied here. 
The normal state behavior in particular exhibits a wide sample-to-sample variation, limiting our ability to meaningfully and quantitatively interpret these measurements at this stage.
The effect on {\Tc}, on the other hand, is remarkably consistent between all measured films (Fig. \ref{fig2}a), which we further parameterize to investigate the superconducting gap symmetry.  
}

\begin{figure}[t]
    \centering
    \includegraphics[width = \columnwidth]{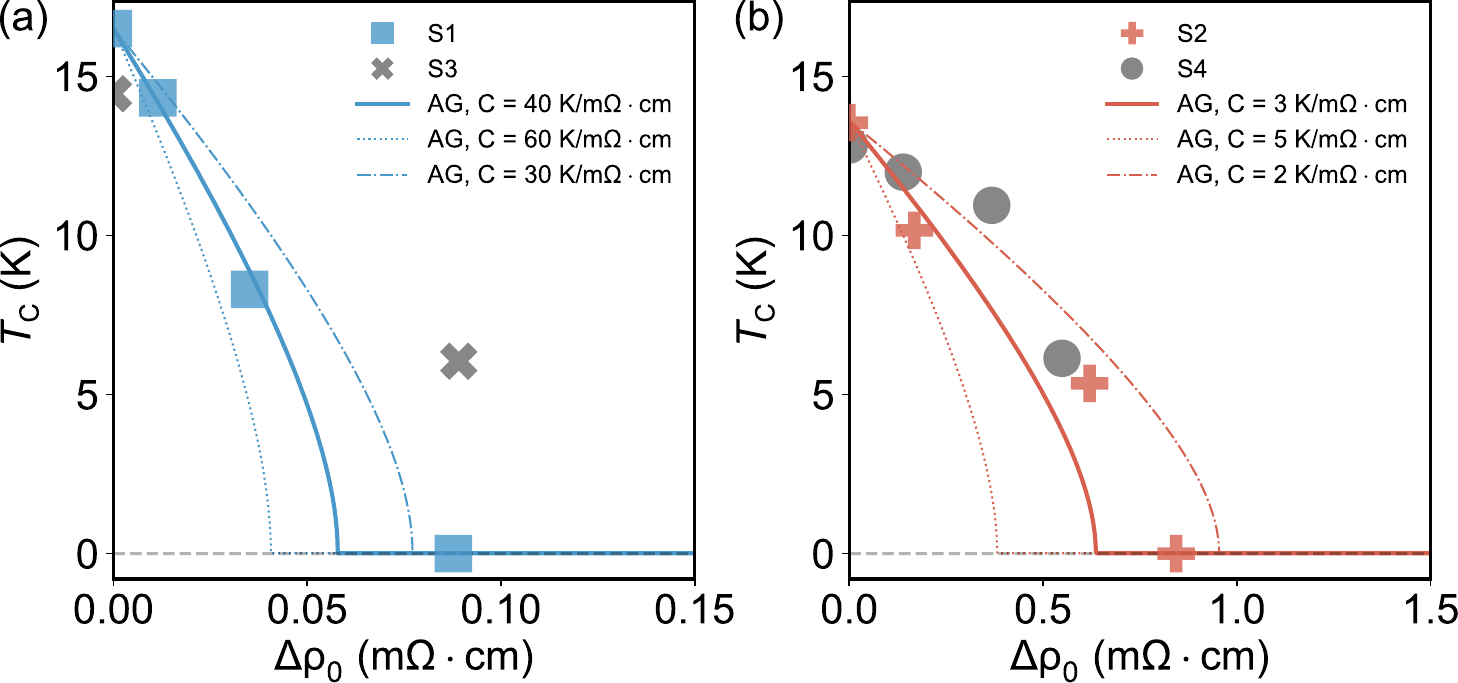}
    \caption{Superconducting transition temperature {\Tc} as a function of change in residual resistivity $\Delta \rho_0$ for (a) S1 and S3, (b) S2 and S4 measured after progressive irradiation doses. The solid lines in blue (red) are generated by fitting the generalized Abrikosov-Gork'kov pair breaking function to data points from S1 (S2) with a scaling parameters $C$ = \textcolor{darkgreen}{40 (3)}  K/m$\Omega \cdot$cm, as described in equation \ref{eq:AG2}. \textcolor{darkgreen}{Additional AG functions (dotted and dashed) corresponding to other $C$ values are plotted for comparison.} The points for S3 and S4 are not included in the fits.}
    \label{fig_ag}
\end{figure}

The sensitivity of unconventional superconductivity to non-magnetic disorder can be described through a generalization of the Abrikosov-Gork'kov (AG) pair-breaking theory for magnetic impurities in a fully gapped superconductor \cite{abrikosov_gorkov, Larkin1965, Millis1988, Radtke1993}. 
While the order parameter for conventional fully gapped $s$-wave superconductivity is largely insensitive to non-magnetic disorder \cite{Anderson1959, Xiao2015_dopedSTO,cavanagh2020_robust_s_wave, Roppongi2023}, an unconventional sign-changing order parameter should be more susceptible to pair-breaking such that increased disorder suppresses superconductivity \cite{legris1993, Rullier2003, Mackenzie_SRO_1998}.
More explicitly, within the AG formulation, the superconducting transition temperature {\Tc} follows
\begin{equation}
    \mathrm{ln}{\left(\dfrac{T_{c0}}{T_c}\right)}= \Psi\left(\frac{1}{2} + \frac{\alpha T_{c0}}{2 \pi T_c} \right) - \Psi\left(\dfrac{1}{2}\right), 
\label{eq:AG}
\end{equation}
where $\Psi$ is the digamma function, the pair-breaking parameter is represented by $\alpha = \hbar / 2 \tau k_B T_{c0}$, $\tau$ is the mean free time due to elastic scattering by impurities, and $T_{c0}$ is {\Tc} in the low-disorder limit ($\alpha \rightarrow 0$). 
In a simple theory, the scattering rate manifests in resistivity as $1/\tau =  (\omega^2_{pl} / 4\pi) \rho_0$ where $\hbar \omega_{pl}$ is the characteristic plasma frequency of in-plane charge carriers.  
\textcolor{darkgreen}{For historical reasons, we note the use of Gaussian CGS units, i.e., the vacuum permittivity $\varepsilon_0 = 1/4\pi$, in this formulation. }
When parameters such as $\tau$ or $\omega_{pl}$ are known for a given material, this relation can -- in principle -- provide quantitative insight to disorder and superconductivity \cite{ruf_noad2024_SRO}. 

Here, we fit the results of our irradiation experiments to a more generalized version of Equation \ref{eq:AG}:
\begin{equation}
    \mathrm{ln}{\left(\dfrac{T_{c0}}{T_c}\right)}= \Psi\left(\frac{1}{2} + C \frac{ \Delta\rho_0 }{ T_c}  \right) - \Psi\left(\dfrac{1}{2}\right), 
\label{eq:AG2}
\end{equation}
where \textcolor{darkgreen}{we use a scaling parameter $C=\omega_{pl}^2(\varepsilon_0 \hbar)/(4 \pi k_B)$  to fit the experimental transport data measured in SI units }\textcolor{darkgreen}{
\footnote{
\textcolor{darkgreen}{
More explicitly, we focus on the second term of the first digamma function argument in Eq. 1 and start by substituting $\alpha = \hbar / 2\tau k_B T_\textrm{c0}$ to rewrite as
\begin{equation}
    \frac{\alpha T_\textrm{c0} }{2 \pi T_\textrm{c}} = \frac{\hbar}{4 \pi k_B T_\textrm{c} \tau}.
\label{eq:sub1}
\end{equation}
Further substituting the Drude expression for scattering time $1/\tau =  (\omega^2_{pl} / 4\pi) \rho_0$ into the same term, we get
\begin{equation}
   \frac{\rho^G_0}{T_\textrm{c}}\frac{\hbar \omega_{pl}^2}{4 \pi k_B *4 \pi}
\label{eq:sub2} 
\end{equation}
where $\rho^G_0$ denotes the resistivity in Gaussian CGS units (dimensions of time). For comparison to our experimental measurements, we convert into SI units by 
\begin{equation}
   \rho^G_0 = \rho^I_0 * 4 \pi \varepsilon_0
\label{eq:CGS} 
\end{equation}
where $\rho^I_0$ is the resistivity in SI units (dimensions of Ohm meters). 
Combining Eq. \ref{eq:sub2} with Eq. \ref{eq:CGS} and taking the assumption $\rho_0 \rightarrow \Delta\rho_0$ as discussed in the text, we obtain a simplified digamma argument term, $C \Delta\rho^I_0 / T_\textrm{c}$, where 
\begin{equation}
    C = \frac{\varepsilon_0 \hbar}{4 \pi k_B}\omega_{pl}^2
\label{eq:Cdef}
\end{equation}
in SI units, with dimensions of Kelvin per resistivity. All resistivities reported within the text are $\rho^I$
}
}
.}
For each sample, we take $T_{c0}$ as the pristine {\Tc} measured prior to irradiation, 16.5 and 13.6 K for S1 and S2 respectively, but note that with the sparsity of data points here the quality of the fit is roughly independent of $T_{c0}$ values within this range.

As shown in Fig. \ref{fig_ag}, the disorder-dependence of both samples S1 and S2 can be fit to this generalized AG model with appropriately chosen $C$ values, here \textcolor{darkgreen}{40 and 3 K/m$\Omega\cdot$cm}, respectively. 
\textcolor{darkgreen}{Additional fits are included for each sample to illustrate the relatively low sensitivity on the precise value of $C$ }
\footnote{
\textcolor{darkgreen}{From optical measurements of Nd$_{0.8}$Sr$_{0.2}$NiO$_2$ films on SrTiO$_3$, Cervasio et al. \cite{cervasio2023optical} reported a room-temperature plasma frequency of $\sim5500$ cm$^{-1}$ ($\hbar \omega_{pl}\sim$0.68 eV), which would correspond to $C\sim57$ K/m$\Omega\cdot$cm in our formulation. 
Encouragingly, this falls within the same order of magnitude as our fits to S1, though we note several challenges for more precise comparison, which include accounting for sample-to-sample variation with the reported transport data, temperature dependence, and low sensitivity of the fit dependence on $C$.}
}.
Samples S3 and S4 are pieces from the same original perovskite film growths as S1 and S2 respectively (i.e., ``sister samples'') which are not included in the AG fits, but here included for comparison to emphasize that both characteristic magnitudes of $\Delta \rho_0$ are reproducible across multiple samples.    

In principle, the critical resistivity at which superconductivity is fully suppressed can be related to other quantitative parameters such as characteristic length scales or critical disorder limits \cite{Mackenzie_SRO_1998, Rullier2003, EmeryKivelson_1995, EmeryKivelson2_1995}.
As in the normal state transport, however, the huge differences in our $C$ scaling factors or critical resistivities for individual samples (note the order-of-magnitude difference in the $x$-axes of Figs. \ref{fig_ag}a,b) suggest that even by accounting for different $\rho_0^\mathrm{pristine}$ through $\Delta \rho_0$, we cannot fully capture varying degrees of non-point-like disorder across different films. 
\footnote{In an ideal unconventional superconductor with only point defects, the critical mean free path $\lambda^{\textrm{crit}}$ for the appearance of superconductivity provides an estimate of the superconducting coherence length $\xi$.  
Notwithstanding the concerns about sample-to-sample variation of $\Delta \rho_0$ for a given radiation dose (Fig. \ref{fig2}b), one reasonable hypothesis would be to consider the lowest $\Delta \rho_0$ values (Fig. \ref{fig_ag}a) as the closest to the intrinsic values that would result from purely point defect scattering. 
It is then possible to estimate $\lambda^{\textrm{crit}}$ from $\Delta \rho_0^{\textrm{crit}}$ $\cong$ 60 {\textmu}$\Omega\cdot$cm \textcolor{darkgreen}{in sample S1}. 
In a sweeping approximation, we use the 2D expression $\lambda\cong (1/\rho)(hd/(e^2 k_F)$ , where $d$ is the interlayer spacing and $k_F$ the Fermi wavevector, taking $k_F\sim$  0.45 Å$^{-1}$ from recent photoemission experiments \cite{ding2024cuprate,Sun2025_LSNO_ARPES}, considering only the large quasi-two dimensional Fermi surface and neglecting the electron pockets near the zone corners.  
The resulting estimate of $\lambda^\textrm{crit}$ $\cong$ $\xi$ $\cong$ 33 Å is therefore only a rough one, but yields a value for $\xi$ in reasonable agreement with estimates from other measurements \cite{li_superconductivity_2019,wang2021isotropic}.}

In addition to the AG model considered above, within the density of our sampling along the disorder parameter, our results also appear generally consistent with other proposed models describing the impact of disorder in superconductors, such as that of phase fluctuations proposed by Emery and Kivelson \cite{EmeryKivelson_1995, EmeryKivelson2_1995} or spatial variance in the order parameter described by Franz, et al. \cite{Franz1997_PRB}, both of which predict systematic suppression of {\Tc} with disorder. 
It is possible that future experiments with more sampling points guided by the universal dose-dependent suppression rate of $T_\textrm{c}$ reported here could be used to more clearly distinguish between such scenarios.

Most generally, the evolution of superconductivity with high-energy electron irradiation allows us to narrow the possible pairing symmetries in superconducting infinite-layer nickelates. 
Our results are inconsistent with the expected progression for a fully-gapped order parameter, in which the superconducting transition temperature would be largely unaffected by irradiation-induced \textcolor{darkgreen}{non-magnetic} disorder \textcolor{darkgreen}{\cite{Blinkin2006_MgB2_s-wave}}.
In the case of an anisotropic $s$-wave order parameter, we would expect an initial disorder-driven decrease in {\Tc} to eventually plateau, never reaching a complete suppression as we observe here \cite{Roppongi2023}. 
Rather, the clear and complete suppression of superconductivity with pair-breaking disorder in NSNO thin films indicates a momentum-space sign-change in the phase of the superconducting order parameter which averages to zero with sufficient scattering (disorder) \cite{Mackenzie2003_rmp_review}, pointing to unconventional superconductivity in infinite-layer nickelates. 
% \textbf{** comment about sign-changing vs "true" nodal ? **}
In future experiments, it may be instructive to consider the wider family of square-planar nickelate superconductors and investigate whether the universality we find in the NSNO films here extends to the Pr- and La-based (Pr/La)$_{1-x}$(Sr/Ca/Eu)$_{x}$NiO$_2$ compounds \cite{osada2020superconducting, osada2021nickelate, zeng2022superconductivity, Wei2023_NENO} as well as the multilayered $T'$ Nd$_{n+1}$Ni$_{n}$O$_{2n+2}$ nickelates \cite{pan2022superconductivity}.
The impact of disorder at different hole dopings maybe also provide some insights to proximal phases on the under- and overdoped sides of the superconducting dome \cite{kluge1995clear, Fukuzumi1996, Li2020_Dome, zeng2020phase, Lee2023_normal_state}.

\vspace{12 pt}

\section{Acknowledgements}
We thank Y. Lee, X. Wei, Y. Mizukami, G.M. Ferguson and J. Sichelschmidt for useful discussions; we thank Ulrich Burkhardt for assistance with sample preparation.
This work was supported by the Max Planck Society.
B.H.G. was supported by Schmidt Science Fellows in partnership with the Rhodes Trust.
Work at SLAC and Stanford was supported by the U. S. Department of Energy, Office of Basic Energy Sciences, Division of Materials Sciences and Engineering (Contract No. DE-AC02-76SF00515).
% \textbf{(please send additions)}. 
The authors acknowledge support from the EMIR\&A French network (FR CNRS 3618) on the platform SIRIUS.
% Electron irradiation was performed under EMIR\&A Project Nos. 24-8965 and 23-2430.

\bibliography{references}

\begin{thebibliography}{76}%
\makeatletter
\providecommand \@ifxundefined [1]{%
 \@ifx{#1\undefined}
}%
\providecommand \@ifnum [1]{%
 \ifnum #1\expandafter \@firstoftwo
 \else \expandafter \@secondoftwo
 \fi
}%
\providecommand \@ifx [1]{%
 \ifx #1\expandafter \@firstoftwo
 \else \expandafter \@secondoftwo
 \fi
}%
\providecommand \natexlab [1]{#1}%
\providecommand \enquote  [1]{``#1''}%
\providecommand \bibnamefont  [1]{#1}%
\providecommand \bibfnamefont [1]{#1}%
\providecommand \citenamefont [1]{#1}%
\providecommand \href@noop [0]{\@secondoftwo}%
\providecommand \href [0]{\begingroup \@sanitize@url \@href}%
\providecommand \@href[1]{\@@startlink{#1}\@@href}%
\providecommand \@@href[1]{\endgroup#1\@@endlink}%
\providecommand \@sanitize@url [0]{\catcode `\\12\catcode `\$12\catcode `\&12\catcode `\#12\catcode `\^12\catcode `\_12\catcode `\%12\relax}%
\providecommand \@@startlink[1]{}%
\providecommand \@@endlink[0]{}%
\providecommand \url  [0]{\begingroup\@sanitize@url \@url }%
\providecommand \@url [1]{\endgroup\@href {#1}{\urlprefix }}%
\providecommand \urlprefix  [0]{URL }%
\providecommand \Eprint [0]{\href }%
\providecommand \doibase [0]{https://doi.org/}%
\providecommand \selectlanguage [0]{\@gobble}%
\providecommand \bibinfo  [0]{\@secondoftwo}%
\providecommand \bibfield  [0]{\@secondoftwo}%
\providecommand \translation [1]{[#1]}%
\providecommand \BibitemOpen [0]{}%
\providecommand \bibitemStop [0]{}%
\providecommand \bibitemNoStop [0]{.\EOS\space}%
\providecommand \EOS [0]{\spacefactor3000\relax}%
\providecommand \BibitemShut  [1]{\csname bibitem#1\endcsname}%
\let\auto@bib@innerbib\@empty
%</preamble>
\bibitem [{\citenamefont {Anisimov}\ \emph {et~al.}(1999)\citenamefont {Anisimov}, \citenamefont {Bukhvalov},\ and\ \citenamefont {Rice}}]{anisimov_1999_prb}%
  \BibitemOpen
  \bibfield  {author} {\bibinfo {author} {\bibfnamefont {V.~I.}\ \bibnamefont {Anisimov}}, \bibinfo {author} {\bibfnamefont {D.}~\bibnamefont {Bukhvalov}},\ and\ \bibinfo {author} {\bibfnamefont {T.~M.}\ \bibnamefont {Rice}},\ }\bibfield  {title} {\bibinfo {title} {Electronic structure of possible nickelate analogs to the cuprates},\ }\href {https://doi.org/10.1103/PhysRevB.59.7901} {\bibfield  {journal} {\bibinfo  {journal} {Phys. Rev. B}\ }\textbf {\bibinfo {volume} {59}},\ \bibinfo {pages} {7901} (\bibinfo {year} {1999})}\BibitemShut {NoStop}%
\bibitem [{\citenamefont {Li}\ \emph {et~al.}(2019)\citenamefont {Li}, \citenamefont {Lee}, \citenamefont {Wang}, \citenamefont {Osada}, \citenamefont {Crossley}, \citenamefont {Lee}, \citenamefont {Cui}, \citenamefont {Hikita},\ and\ \citenamefont {Hwang}}]{li_superconductivity_2019}%
  \BibitemOpen
  \bibfield  {author} {\bibinfo {author} {\bibfnamefont {D.}~\bibnamefont {Li}}, \bibinfo {author} {\bibfnamefont {K.}~\bibnamefont {Lee}}, \bibinfo {author} {\bibfnamefont {B.~Y.}\ \bibnamefont {Wang}}, \bibinfo {author} {\bibfnamefont {M.}~\bibnamefont {Osada}}, \bibinfo {author} {\bibfnamefont {S.}~\bibnamefont {Crossley}}, \bibinfo {author} {\bibfnamefont {H.~R.}\ \bibnamefont {Lee}}, \bibinfo {author} {\bibfnamefont {Y.}~\bibnamefont {Cui}}, \bibinfo {author} {\bibfnamefont {Y.}~\bibnamefont {Hikita}},\ and\ \bibinfo {author} {\bibfnamefont {H.~Y.}\ \bibnamefont {Hwang}},\ }\bibfield  {title} {\bibinfo {title} {Superconductivity in an infinite-layer nickelate},\ }\href {https://doi.org/10.1038/s41586-019-1496-5} {\bibfield  {journal} {\bibinfo  {journal} {Nature}\ }\textbf {\bibinfo {volume} {572}},\ \bibinfo {pages} {624} (\bibinfo {year} {2019})}\BibitemShut {NoStop}%
\bibitem [{\citenamefont {Bednorz}\ and\ \citenamefont {Müller}(1986)}]{Bednorz1986_cuprate}%
  \BibitemOpen
  \bibfield  {author} {\bibinfo {author} {\bibfnamefont {J.~G.}\ \bibnamefont {Bednorz}}\ and\ \bibinfo {author} {\bibfnamefont {K.~A.}\ \bibnamefont {Müller}},\ }\bibfield  {title} {\bibinfo {title} {{Possible high $T_\mathrm{c}$ superconductivity in the B-La-Cu-O system}},\ }\href {https://doi.org/10.1007/bf01303701} {\bibfield  {journal} {\bibinfo  {journal} {Zeitschrift für Physik B Condensed Matter}\ }\textbf {\bibinfo {volume} {64}},\ \bibinfo {pages} {189–193} (\bibinfo {year} {1986})}\BibitemShut {NoStop}%
\bibitem [{\citenamefont {Wang}\ \emph {et~al.}(2024)\citenamefont {Wang}, \citenamefont {Lee},\ and\ \citenamefont {Goodge}}]{WangLeeGoodge-Review}%
  \BibitemOpen
  \bibfield  {author} {\bibinfo {author} {\bibfnamefont {B.~Y.}\ \bibnamefont {Wang}}, \bibinfo {author} {\bibfnamefont {K.}~\bibnamefont {Lee}},\ and\ \bibinfo {author} {\bibfnamefont {B.~H.}\ \bibnamefont {Goodge}},\ }\bibfield  {title} {\bibinfo {title} {{Experimental Progress in Superconducting Nickelates}},\ }\href {https://doi.org/https://doi.org/10.1146/annurev-conmatphys-032922-093307} {\bibfield  {journal} {\bibinfo  {journal} {Annu. Rev. Condens. Matter Phys.}\ }\textbf {\bibinfo {volume} {15}},\ \bibinfo {pages} {305} (\bibinfo {year} {2024})}\BibitemShut {NoStop}%
\bibitem [{\citenamefont {Lee}\ and\ \citenamefont {Pickett}(2004)}]{Lee2004_Ni_is_not_Cu}%
  \BibitemOpen
  \bibfield  {author} {\bibinfo {author} {\bibfnamefont {K.-W.}\ \bibnamefont {Lee}}\ and\ \bibinfo {author} {\bibfnamefont {W.~E.}\ \bibnamefont {Pickett}},\ }\bibfield  {title} {\bibinfo {title} {{Infinite-layer $\mathrm{La}\mathrm{Ni}{\mathrm{O}}_{2}$: ${\mathrm{Ni}}^{1+}$ is not ${\mathrm{Cu}}^{2+}$}},\ }\href {https://doi.org/10.1103/PhysRevB.70.165109} {\bibfield  {journal} {\bibinfo  {journal} {Phys. Rev. B}\ }\textbf {\bibinfo {volume} {70}},\ \bibinfo {pages} {165109} (\bibinfo {year} {2004})}\BibitemShut {NoStop}%
\bibitem [{\citenamefont {Hepting}\ \emph {et~al.}(2020)\citenamefont {Hepting}, \citenamefont {Li}, \citenamefont {Jia}, \citenamefont {Lu}, \citenamefont {Paris}, \citenamefont {Tseng}, \citenamefont {Feng}, \citenamefont {Osada}, \citenamefont {Been}, \citenamefont {Hikita}, \citenamefont {Chuang}, \citenamefont {Hussain}, \citenamefont {Zhou}, \citenamefont {Nag}, \citenamefont {Garcia-Fernandez}, \citenamefont {Rossi}, \citenamefont {Huang}, \citenamefont {Huang}, \citenamefont {Shen}, \citenamefont {Schmitt}, \citenamefont {Hwang}, \citenamefont {Moritz}, \citenamefont {Zaanen}, \citenamefont {Devereaux},\ and\ \citenamefont {Lee}}]{Hepting2020_parent_electronic_structure}%
  \BibitemOpen
  \bibfield  {author} {\bibinfo {author} {\bibfnamefont {M.}~\bibnamefont {Hepting}}, \bibinfo {author} {\bibfnamefont {D.}~\bibnamefont {Li}}, \bibinfo {author} {\bibfnamefont {C.~J.}\ \bibnamefont {Jia}}, \bibinfo {author} {\bibfnamefont {H.}~\bibnamefont {Lu}}, \bibinfo {author} {\bibfnamefont {E.}~\bibnamefont {Paris}}, \bibinfo {author} {\bibfnamefont {Y.}~\bibnamefont {Tseng}}, \bibinfo {author} {\bibfnamefont {X.}~\bibnamefont {Feng}}, \bibinfo {author} {\bibfnamefont {M.}~\bibnamefont {Osada}}, \bibinfo {author} {\bibfnamefont {E.}~\bibnamefont {Been}}, \bibinfo {author} {\bibfnamefont {Y.}~\bibnamefont {Hikita}}, \bibinfo {author} {\bibfnamefont {Y.-D.}\ \bibnamefont {Chuang}}, \bibinfo {author} {\bibfnamefont {Z.}~\bibnamefont {Hussain}}, \bibinfo {author} {\bibfnamefont {K.~J.}\ \bibnamefont {Zhou}}, \bibinfo {author} {\bibfnamefont {A.}~\bibnamefont {Nag}}, \bibinfo {author} {\bibfnamefont {M.}~\bibnamefont {Garcia-Fernandez}}, \bibinfo {author} {\bibfnamefont {M.}~\bibnamefont {Rossi}}, \bibinfo
  {author} {\bibfnamefont {H.~Y.}\ \bibnamefont {Huang}}, \bibinfo {author} {\bibfnamefont {D.~J.}\ \bibnamefont {Huang}}, \bibinfo {author} {\bibfnamefont {Z.~X.}\ \bibnamefont {Shen}}, \bibinfo {author} {\bibfnamefont {T.}~\bibnamefont {Schmitt}}, \bibinfo {author} {\bibfnamefont {H.~Y.}\ \bibnamefont {Hwang}}, \bibinfo {author} {\bibfnamefont {B.}~\bibnamefont {Moritz}}, \bibinfo {author} {\bibfnamefont {J.}~\bibnamefont {Zaanen}}, \bibinfo {author} {\bibfnamefont {T.~P.}\ \bibnamefont {Devereaux}},\ and\ \bibinfo {author} {\bibfnamefont {W.~S.}\ \bibnamefont {Lee}},\ }\bibfield  {title} {\bibinfo {title} {Electronic structure of the parent compound of superconducting infinite-layer nickelates},\ }\href {https://doi.org/10.1038/s41563-019-0585-z} {\bibfield  {journal} {\bibinfo  {journal} {Nat. Mater.}\ }\textbf {\bibinfo {volume} {19}},\ \bibinfo {pages} {381–385} (\bibinfo {year} {2020})}\BibitemShut {NoStop}%
\bibitem [{\citenamefont {Goodge}\ \emph {et~al.}(2021)\citenamefont {Goodge}, \citenamefont {Li}, \citenamefont {Lee}, \citenamefont {Osada}, \citenamefont {Wang}, \citenamefont {Sawatzky}, \citenamefont {Hwang},\ and\ \citenamefont {Kourkoutis}}]{goodge2021doping}%
  \BibitemOpen
  \bibfield  {author} {\bibinfo {author} {\bibfnamefont {B.~H.}\ \bibnamefont {Goodge}}, \bibinfo {author} {\bibfnamefont {D.}~\bibnamefont {Li}}, \bibinfo {author} {\bibfnamefont {K.}~\bibnamefont {Lee}}, \bibinfo {author} {\bibfnamefont {M.}~\bibnamefont {Osada}}, \bibinfo {author} {\bibfnamefont {B.~Y.}\ \bibnamefont {Wang}}, \bibinfo {author} {\bibfnamefont {G.~A.}\ \bibnamefont {Sawatzky}}, \bibinfo {author} {\bibfnamefont {H.~Y.}\ \bibnamefont {Hwang}},\ and\ \bibinfo {author} {\bibfnamefont {L.~F.}\ \bibnamefont {Kourkoutis}},\ }\bibfield  {title} {\bibinfo {title} {Doping evolution of the {M}ott--{H}ubbard landscape in infinite-layer nickelates},\ }\href {http://dx.doi.org/10.1073/pnas.2007683118} {\bibfield  {journal} {\bibinfo  {journal} {Proceedings of the National Academy of Sciences}\ }\textbf {\bibinfo {volume} {118}},\ \bibinfo {pages} {e2007683118} (\bibinfo {year} {2021})}\BibitemShut {NoStop}%
\bibitem [{\citenamefont {Gu}\ \emph {et~al.}(2020)\citenamefont {Gu}, \citenamefont {Li}, \citenamefont {Wan}, \citenamefont {Li}, \citenamefont {Guo}, \citenamefont {Yang}, \citenamefont {Li}, \citenamefont {Zhu}, \citenamefont {Pan}, \citenamefont {Nie},\ and\ \citenamefont {Wen}}]{gu2020single}%
  \BibitemOpen
  \bibfield  {author} {\bibinfo {author} {\bibfnamefont {Q.}~\bibnamefont {Gu}}, \bibinfo {author} {\bibfnamefont {Y.}~\bibnamefont {Li}}, \bibinfo {author} {\bibfnamefont {S.}~\bibnamefont {Wan}}, \bibinfo {author} {\bibfnamefont {H.}~\bibnamefont {Li}}, \bibinfo {author} {\bibfnamefont {W.}~\bibnamefont {Guo}}, \bibinfo {author} {\bibfnamefont {H.}~\bibnamefont {Yang}}, \bibinfo {author} {\bibfnamefont {Q.}~\bibnamefont {Li}}, \bibinfo {author} {\bibfnamefont {X.}~\bibnamefont {Zhu}}, \bibinfo {author} {\bibfnamefont {X.}~\bibnamefont {Pan}}, \bibinfo {author} {\bibfnamefont {Y.}~\bibnamefont {Nie}},\ and\ \bibinfo {author} {\bibfnamefont {H.-H.}\ \bibnamefont {Wen}},\ }\bibfield  {title} {\bibinfo {title} {{Single particle tunneling spectrum of superconducting Nd$_{1-x}$Sr$_x$NiO$_2$ thin films}},\ }\href {https://doi.org/10.1038/s41467-020-19908-1} {\bibfield  {journal} {\bibinfo  {journal} {Nat Commun}\ }\textbf {\bibinfo {volume} {11}},\ \bibinfo {pages} {6027} (\bibinfo {year} {2020})}\BibitemShut
  {NoStop}%
\bibitem [{\citenamefont {Harvey}\ \emph {et~al.}(2022)\citenamefont {Harvey}, \citenamefont {Wang}, \citenamefont {Fowlie}, \citenamefont {Osada}, \citenamefont {Lee}, \citenamefont {Lee}, \citenamefont {Li},\ and\ \citenamefont {Hwang}}]{Harvey2022MutualInductance}%
  \BibitemOpen
  \bibfield  {author} {\bibinfo {author} {\bibfnamefont {S.~P.}\ \bibnamefont {Harvey}}, \bibinfo {author} {\bibfnamefont {B.~Y.}\ \bibnamefont {Wang}}, \bibinfo {author} {\bibfnamefont {J.}~\bibnamefont {Fowlie}}, \bibinfo {author} {\bibfnamefont {M.}~\bibnamefont {Osada}}, \bibinfo {author} {\bibfnamefont {K.}~\bibnamefont {Lee}}, \bibinfo {author} {\bibfnamefont {Y.}~\bibnamefont {Lee}}, \bibinfo {author} {\bibfnamefont {D.}~\bibnamefont {Li}},\ and\ \bibinfo {author} {\bibfnamefont {H.~Y.}\ \bibnamefont {Hwang}},\ }\bibfield  {title} {\bibinfo {title} {Evidence for nodal superconductivity in infinite-layer nickelates},\ }\href {httpfss://doi.org/10.48550/arXiv.2201.12971} {\bibfield  {journal} {\bibinfo  {journal} {arXiv:2201.12971}\ } (\bibinfo {year} {2022})}\BibitemShut {NoStop}%
\bibitem [{\citenamefont {Chow}\ \emph {et~al.}(2023)\citenamefont {Chow}, \citenamefont {Sudheesh}, \citenamefont {Luo}, \citenamefont {Nandi}, \citenamefont {Heil}, \citenamefont {Deuschle}, \citenamefont {Zeng}, \citenamefont {Zhang}, \citenamefont {Prakash}, \citenamefont {Du}, \citenamefont {Lim}, \citenamefont {van Aken}, \citenamefont {Chia},\ and\ \citenamefont {Ariando}}]{chow2023pairing}%
  \BibitemOpen
  \bibfield  {author} {\bibinfo {author} {\bibfnamefont {L.~E.}\ \bibnamefont {Chow}}, \bibinfo {author} {\bibfnamefont {S.~K.}\ \bibnamefont {Sudheesh}}, \bibinfo {author} {\bibfnamefont {Z.~Y.}\ \bibnamefont {Luo}}, \bibinfo {author} {\bibfnamefont {P.}~\bibnamefont {Nandi}}, \bibinfo {author} {\bibfnamefont {T.}~\bibnamefont {Heil}}, \bibinfo {author} {\bibfnamefont {J.}~\bibnamefont {Deuschle}}, \bibinfo {author} {\bibfnamefont {S.~W.}\ \bibnamefont {Zeng}}, \bibinfo {author} {\bibfnamefont {Z.~T.}\ \bibnamefont {Zhang}}, \bibinfo {author} {\bibfnamefont {S.}~\bibnamefont {Prakash}}, \bibinfo {author} {\bibfnamefont {X.~M.}\ \bibnamefont {Du}}, \bibinfo {author} {\bibfnamefont {Z.~S.}\ \bibnamefont {Lim}}, \bibinfo {author} {\bibfnamefont {P.~A.}\ \bibnamefont {van Aken}}, \bibinfo {author} {\bibfnamefont {E.~E.~M.}\ \bibnamefont {Chia}},\ and\ \bibinfo {author} {\bibfnamefont {A.}~\bibnamefont {Ariando}},\ }\bibfield  {title} {\bibinfo {title} {Pairing symmetry in infinite-layer nickelate
  superconductor},\ }\href {https://arxiv.org/abs/2201.10038} {\bibfield  {journal} {\bibinfo  {journal} {arXiv:2201.10038}\ } (\bibinfo {year} {2023})}\BibitemShut {NoStop}%
\bibitem [{\citenamefont {Cheng}\ \emph {et~al.}(2024)\citenamefont {Cheng}, \citenamefont {Cheng}, \citenamefont {Lee}, \citenamefont {Luo}, \citenamefont {Chen}, \citenamefont {Lee}, \citenamefont {Wang}, \citenamefont {Mootz}, \citenamefont {Perakis}, \citenamefont {Shen}, \citenamefont {Hwang},\ and\ \citenamefont {Wang}}]{cheng2024evidence}%
  \BibitemOpen
  \bibfield  {author} {\bibinfo {author} {\bibfnamefont {B.}~\bibnamefont {Cheng}}, \bibinfo {author} {\bibfnamefont {D.}~\bibnamefont {Cheng}}, \bibinfo {author} {\bibfnamefont {K.}~\bibnamefont {Lee}}, \bibinfo {author} {\bibfnamefont {L.}~\bibnamefont {Luo}}, \bibinfo {author} {\bibfnamefont {Z.}~\bibnamefont {Chen}}, \bibinfo {author} {\bibfnamefont {Y.}~\bibnamefont {Lee}}, \bibinfo {author} {\bibfnamefont {B.~Y.}\ \bibnamefont {Wang}}, \bibinfo {author} {\bibfnamefont {M.}~\bibnamefont {Mootz}}, \bibinfo {author} {\bibfnamefont {I.~E.}\ \bibnamefont {Perakis}}, \bibinfo {author} {\bibfnamefont {Z.-X.}\ \bibnamefont {Shen}}, \bibinfo {author} {\bibfnamefont {H.~Y.}\ \bibnamefont {Hwang}},\ and\ \bibinfo {author} {\bibfnamefont {J.}~\bibnamefont {Wang}},\ }\bibfield  {title} {\bibinfo {title} {Evidence for d-wave superconductivity of infinite-layer nickelates from low-energy electrodynamics},\ }\href {https://doi.org/10.1038/s41563-023-01766-z} {\bibfield  {journal} {\bibinfo  {journal} {Nat. Mater.}\
  }\textbf {\bibinfo {volume} {23}},\ \bibinfo {pages} {775} (\bibinfo {year} {2024})}\BibitemShut {NoStop}%
\bibitem [{\citenamefont {Grissonnanche}\ \emph {et~al.}(2024)\citenamefont {Grissonnanche}, \citenamefont {Pan}, \citenamefont {LaBollita}, \citenamefont {Segedin}, \citenamefont {Song}, \citenamefont {Paik}, \citenamefont {Brooks}, \citenamefont {Beauchesne-Blanchet}, \citenamefont {Santana~Gonz\'alez}, \citenamefont {Botana}, \citenamefont {Mundy},\ and\ \citenamefont {Ramshaw}}]{Grissonnanche2024_seebeck}%
  \BibitemOpen
  \bibfield  {author} {\bibinfo {author} {\bibfnamefont {G.}~\bibnamefont {Grissonnanche}}, \bibinfo {author} {\bibfnamefont {G.~A.}\ \bibnamefont {Pan}}, \bibinfo {author} {\bibfnamefont {H.}~\bibnamefont {LaBollita}}, \bibinfo {author} {\bibfnamefont {D.~F.}\ \bibnamefont {Segedin}}, \bibinfo {author} {\bibfnamefont {Q.}~\bibnamefont {Song}}, \bibinfo {author} {\bibfnamefont {H.}~\bibnamefont {Paik}}, \bibinfo {author} {\bibfnamefont {C.~M.}\ \bibnamefont {Brooks}}, \bibinfo {author} {\bibfnamefont {E.}~\bibnamefont {Beauchesne-Blanchet}}, \bibinfo {author} {\bibfnamefont {J.~L.}\ \bibnamefont {Santana~Gonz\'alez}}, \bibinfo {author} {\bibfnamefont {A.~S.}\ \bibnamefont {Botana}}, \bibinfo {author} {\bibfnamefont {J.~A.}\ \bibnamefont {Mundy}},\ and\ \bibinfo {author} {\bibfnamefont {B.~J.}\ \bibnamefont {Ramshaw}},\ }\bibfield  {title} {\bibinfo {title} {Electronic band structure of a superconducting nickelate probed by the seebeck coefficient in the disordered limit},\ }\href
  {https://doi.org/10.1103/PhysRevX.14.041021} {\bibfield  {journal} {\bibinfo  {journal} {Phys. Rev. X}\ }\textbf {\bibinfo {volume} {14}},\ \bibinfo {pages} {041021} (\bibinfo {year} {2024})}\BibitemShut {NoStop}%
\bibitem [{\citenamefont {Abrikosov}\ and\ \citenamefont {Gor'kov}(1961)}]{abrikosov_gorkov}%
  \BibitemOpen
  \bibfield  {author} {\bibinfo {author} {\bibfnamefont {A.}~\bibnamefont {Abrikosov}}\ and\ \bibinfo {author} {\bibfnamefont {L.}~\bibnamefont {Gor'kov}},\ }\bibfield  {title} {\bibinfo {title} {{Contribution} to the theory of superconducting alloys with paramagnetic impurities},\ }\href@noop {} {\bibfield  {journal} {\bibinfo  {journal} {Sov. Phys. JETP}\ }\textbf {\bibinfo {volume} {12}},\ \bibinfo {pages} {1243} (\bibinfo {year} {1961})}\BibitemShut {NoStop}%
\bibitem [{\citenamefont {Radtke}\ \emph {et~al.}(1993)\citenamefont {Radtke}, \citenamefont {Levin}, \citenamefont {Sch\"uttler},\ and\ \citenamefont {Norman}}]{Radtke1993}%
  \BibitemOpen
  \bibfield  {author} {\bibinfo {author} {\bibfnamefont {R.~J.}\ \bibnamefont {Radtke}}, \bibinfo {author} {\bibfnamefont {K.}~\bibnamefont {Levin}}, \bibinfo {author} {\bibfnamefont {H.-B.}\ \bibnamefont {Sch\"uttler}},\ and\ \bibinfo {author} {\bibfnamefont {M.~R.}\ \bibnamefont {Norman}},\ }\bibfield  {title} {\bibinfo {title} {{Predictions for impurity-induced ${\mathit{T}}_{\mathit{c}}$ suppression in the high-temperature superconductors}},\ }\href {https://doi.org/10.1103/PhysRevB.48.653} {\bibfield  {journal} {\bibinfo  {journal} {Phys. Rev. B}\ }\textbf {\bibinfo {volume} {48}},\ \bibinfo {pages} {653} (\bibinfo {year} {1993})}\BibitemShut {NoStop}%
\bibitem [{\citenamefont {Mackenzie}\ and\ \citenamefont {Maeno}(2003)}]{Mackenzie2003_rmp_review}%
  \BibitemOpen
  \bibfield  {author} {\bibinfo {author} {\bibfnamefont {A.~P.}\ \bibnamefont {Mackenzie}}\ and\ \bibinfo {author} {\bibfnamefont {Y.}~\bibnamefont {Maeno}},\ }\bibfield  {title} {\bibinfo {title} {The superconductivity of {S}r$_2${R}u{O}$_4$ and the physics of spin-triplet pairing},\ }\href {https://doi.org/10.1103/RevModPhys.75.657} {\bibfield  {journal} {\bibinfo  {journal} {Rev. Mod. Phys.}\ }\textbf {\bibinfo {volume} {75}},\ \bibinfo {pages} {657} (\bibinfo {year} {2003})}\BibitemShut {NoStop}%
\bibitem [{\citenamefont {Alloul}\ \emph {et~al.}(1991)\citenamefont {Alloul}, \citenamefont {Mendels}, \citenamefont {Casalta}, \citenamefont {Marucco},\ and\ \citenamefont {Arabski}}]{Alloul_YBCO_Zn_1991}%
  \BibitemOpen
  \bibfield  {author} {\bibinfo {author} {\bibfnamefont {H.}~\bibnamefont {Alloul}}, \bibinfo {author} {\bibfnamefont {P.}~\bibnamefont {Mendels}}, \bibinfo {author} {\bibfnamefont {H.}~\bibnamefont {Casalta}}, \bibinfo {author} {\bibfnamefont {J.~F.}\ \bibnamefont {Marucco}},\ and\ \bibinfo {author} {\bibfnamefont {J.}~\bibnamefont {Arabski}},\ }\bibfield  {title} {\bibinfo {title} {Correlations between magnetic and superconducting properties of {Z}n-substituted \ce{YBa2Cu3O6}$_{+x}$},\ }\href {https://doi.org/10.1103/PhysRevLett.67.3140} {\bibfield  {journal} {\bibinfo  {journal} {Phys. Rev. Lett.}\ }\textbf {\bibinfo {volume} {67}},\ \bibinfo {pages} {3140} (\bibinfo {year} {1991})}\BibitemShut {NoStop}%
\bibitem [{\citenamefont {Fukuzumi}\ \emph {et~al.}(1996)\citenamefont {Fukuzumi}, \citenamefont {Mizuhashi}, \citenamefont {Takenaka},\ and\ \citenamefont {Uchida}}]{Fukuzumi1996}%
  \BibitemOpen
  \bibfield  {author} {\bibinfo {author} {\bibfnamefont {Y.}~\bibnamefont {Fukuzumi}}, \bibinfo {author} {\bibfnamefont {K.}~\bibnamefont {Mizuhashi}}, \bibinfo {author} {\bibfnamefont {K.}~\bibnamefont {Takenaka}},\ and\ \bibinfo {author} {\bibfnamefont {S.}~\bibnamefont {Uchida}},\ }\bibfield  {title} {\bibinfo {title} {{Universal Superconductor-Insulator Transition and ${T}_{c}$ Depression in Zn-Substituted High- ${T}_{c}$ Cuprates in the Underdoped Regime}},\ }\href {https://doi.org/10.1103/PhysRevLett.76.684} {\bibfield  {journal} {\bibinfo  {journal} {Phys. Rev. Lett.}\ }\textbf {\bibinfo {volume} {76}},\ \bibinfo {pages} {684} (\bibinfo {year} {1996})}\BibitemShut {NoStop}%
\bibitem [{\citenamefont {Karpi\ifmmode~\acute{n}\else \'{n}\fi{}ska}\ \emph {et~al.}(2000)\citenamefont {Karpi\ifmmode~\acute{n}\else \'{n}\fi{}ska}, \citenamefont {Cieplak}, \citenamefont {Guha}, \citenamefont {Malinowski}, \citenamefont {Sko\ifmmode~\acute{s}\else \'{s}\fi{}kiewicz}, \citenamefont {Plesiewicz}, \citenamefont {Berkowski}, \citenamefont {Boyce}, \citenamefont {Lemberger},\ and\ \citenamefont {Lindenfeld}}]{Karpinska_LSCO_Zn_2000}%
  \BibitemOpen
  \bibfield  {author} {\bibinfo {author} {\bibfnamefont {K.}~\bibnamefont {Karpi\ifmmode~\acute{n}\else \'{n}\fi{}ska}}, \bibinfo {author} {\bibfnamefont {M.~Z.}\ \bibnamefont {Cieplak}}, \bibinfo {author} {\bibfnamefont {S.}~\bibnamefont {Guha}}, \bibinfo {author} {\bibfnamefont {A.}~\bibnamefont {Malinowski}}, \bibinfo {author} {\bibfnamefont {T.}~\bibnamefont {Sko\ifmmode~\acute{s}\else \'{s}\fi{}kiewicz}}, \bibinfo {author} {\bibfnamefont {W.}~\bibnamefont {Plesiewicz}}, \bibinfo {author} {\bibfnamefont {M.}~\bibnamefont {Berkowski}}, \bibinfo {author} {\bibfnamefont {B.}~\bibnamefont {Boyce}}, \bibinfo {author} {\bibfnamefont {T.~R.}\ \bibnamefont {Lemberger}},\ and\ \bibinfo {author} {\bibfnamefont {P.}~\bibnamefont {Lindenfeld}},\ }\bibfield  {title} {\bibinfo {title} {{Metallic Nonsuperconducting Phase and $\mathit{d}$-Wave Superconductivity in Zn-Substituted {La$_{1.85}$Sr$_{0.15}$CuO$_4$}}},\ }\href {https://doi.org/10.1103/PhysRevLett.84.155} {\bibfield  {journal} {\bibinfo  {journal} {Phys. Rev.
  Lett.}\ }\textbf {\bibinfo {volume} {84}},\ \bibinfo {pages} {155} (\bibinfo {year} {2000})}\BibitemShut {NoStop}%
\bibitem [{\citenamefont {Mackenzie}\ \emph {et~al.}(1998)\citenamefont {Mackenzie}, \citenamefont {Haselwimmer}, \citenamefont {Tyler}, \citenamefont {Lonzarich}, \citenamefont {Mori}, \citenamefont {Nishizaki},\ and\ \citenamefont {Maeno}}]{Mackenzie_SRO_1998}%
  \BibitemOpen
  \bibfield  {author} {\bibinfo {author} {\bibfnamefont {A.~P.}\ \bibnamefont {Mackenzie}}, \bibinfo {author} {\bibfnamefont {R.~K.~W.}\ \bibnamefont {Haselwimmer}}, \bibinfo {author} {\bibfnamefont {A.~W.}\ \bibnamefont {Tyler}}, \bibinfo {author} {\bibfnamefont {G.~G.}\ \bibnamefont {Lonzarich}}, \bibinfo {author} {\bibfnamefont {Y.}~\bibnamefont {Mori}}, \bibinfo {author} {\bibfnamefont {S.}~\bibnamefont {Nishizaki}},\ and\ \bibinfo {author} {\bibfnamefont {Y.}~\bibnamefont {Maeno}},\ }\bibfield  {title} {\bibinfo {title} {{Extremely Strong Dependence of Superconductivity on Disorder in \ce{Sr2RuO4}}},\ }\href {https://doi.org/10.1103/PhysRevLett.80.161} {\bibfield  {journal} {\bibinfo  {journal} {Phys. Rev. Lett.}\ }\textbf {\bibinfo {volume} {80}},\ \bibinfo {pages} {161} (\bibinfo {year} {1998})}\BibitemShut {NoStop}%
\bibitem [{\citenamefont {Giapintzakis}\ \emph {et~al.}(1992)\citenamefont {Giapintzakis}, \citenamefont {Lee}, \citenamefont {Rice}, \citenamefont {Ginsberg}, \citenamefont {Robertson}, \citenamefont {Wheeler}, \citenamefont {Kirk},\ and\ \citenamefont {Ruault}}]{Giapintzakis1992_YCBO_irradiation}%
  \BibitemOpen
  \bibfield  {author} {\bibinfo {author} {\bibfnamefont {J.}~\bibnamefont {Giapintzakis}}, \bibinfo {author} {\bibfnamefont {W.~C.}\ \bibnamefont {Lee}}, \bibinfo {author} {\bibfnamefont {J.~P.}\ \bibnamefont {Rice}}, \bibinfo {author} {\bibfnamefont {D.~M.}\ \bibnamefont {Ginsberg}}, \bibinfo {author} {\bibfnamefont {I.~M.}\ \bibnamefont {Robertson}}, \bibinfo {author} {\bibfnamefont {R.}~\bibnamefont {Wheeler}}, \bibinfo {author} {\bibfnamefont {M.~A.}\ \bibnamefont {Kirk}},\ and\ \bibinfo {author} {\bibfnamefont {M.-O.}\ \bibnamefont {Ruault}},\ }\bibfield  {title} {\bibinfo {title} {{Production and identification of flux-pinning defects by electron irradiation in \ce{YBa2Cu3O7}$_{-x}$ single crystals}},\ }\href {https://doi.org/10.1103/PhysRevB.45.10677} {\bibfield  {journal} {\bibinfo  {journal} {Phys. Rev. B}\ }\textbf {\bibinfo {volume} {45}},\ \bibinfo {pages} {10677} (\bibinfo {year} {1992})}\BibitemShut {NoStop}%
\bibitem [{\citenamefont {Konczykowski}\ and\ \citenamefont {Gilchrist}(1991)}]{konczykowskiYBCO_1991}%
  \BibitemOpen
  \bibfield  {author} {\bibinfo {author} {\bibfnamefont {M.}~\bibnamefont {Konczykowski}}\ and\ \bibinfo {author} {\bibfnamefont {J.}~\bibnamefont {Gilchrist}},\ }\bibfield  {title} {\bibinfo {title} {{Electron irradiation of {YBa}$_2${Cu}$_3${O}$_7$ ceramics}},\ }\href {https://doi.org/10.1051/jp3:1991230} {\bibfield  {journal} {\bibinfo  {journal} {{Journal de Physique III}}\ }\textbf {\bibinfo {volume} {1}},\ \bibinfo {pages} {1765} (\bibinfo {year} {1991})}\BibitemShut {NoStop}%
\bibitem [{\citenamefont {Alessi}\ \emph {et~al.}(2023)\citenamefont {Alessi}, \citenamefont {Cavani}, \citenamefont {Grasset}, \citenamefont {Drouhin}, \citenamefont {Safarov},\ and\ \citenamefont {Konczykowski}}]{Alessi2023_Electron_irradiation}%
  \BibitemOpen
  \bibfield  {author} {\bibinfo {author} {\bibfnamefont {A.}~\bibnamefont {Alessi}}, \bibinfo {author} {\bibfnamefont {O.}~\bibnamefont {Cavani}}, \bibinfo {author} {\bibfnamefont {R.}~\bibnamefont {Grasset}}, \bibinfo {author} {\bibfnamefont {H.}~\bibnamefont {Drouhin}}, \bibinfo {author} {\bibfnamefont {V.~I.}\ \bibnamefont {Safarov}},\ and\ \bibinfo {author} {\bibfnamefont {M.}~\bibnamefont {Konczykowski}},\ }\bibfield  {title} {\bibinfo {title} {Electron irradiation: From test to material tailoring},\ }\href {https://iopscience.iop.org/article/10.1209/0295-5075/acf47c/meta} {\bibfield  {journal} {\bibinfo  {journal} {Europhysics Letters}\ }\textbf {\bibinfo {volume} {143}},\ \bibinfo {pages} {56001} (\bibinfo {year} {2023})}\BibitemShut {NoStop}%
\bibitem [{\citenamefont {Sunko}\ \emph {et~al.}(2020)\citenamefont {Sunko}, \citenamefont {McGuinness}, \citenamefont {Chang}, \citenamefont {Zhakina}, \citenamefont {Khim}, \citenamefont {Dreyer}, \citenamefont {Konczykowski}, \citenamefont {Borrmann}, \citenamefont {Moll}, \citenamefont {K\"onig}, \citenamefont {Muller},\ and\ \citenamefont {Mackenzie}}]{sunko2020controlled}%
  \BibitemOpen
  \bibfield  {author} {\bibinfo {author} {\bibfnamefont {V.}~\bibnamefont {Sunko}}, \bibinfo {author} {\bibfnamefont {P.~H.}\ \bibnamefont {McGuinness}}, \bibinfo {author} {\bibfnamefont {C.~S.}\ \bibnamefont {Chang}}, \bibinfo {author} {\bibfnamefont {E.}~\bibnamefont {Zhakina}}, \bibinfo {author} {\bibfnamefont {S.}~\bibnamefont {Khim}}, \bibinfo {author} {\bibfnamefont {C.~E.}\ \bibnamefont {Dreyer}}, \bibinfo {author} {\bibfnamefont {M.}~\bibnamefont {Konczykowski}}, \bibinfo {author} {\bibfnamefont {H.}~\bibnamefont {Borrmann}}, \bibinfo {author} {\bibfnamefont {P.~J.~W.}\ \bibnamefont {Moll}}, \bibinfo {author} {\bibfnamefont {M.}~\bibnamefont {K\"onig}}, \bibinfo {author} {\bibfnamefont {D.~A.}\ \bibnamefont {Muller}},\ and\ \bibinfo {author} {\bibfnamefont {A.~P.}\ \bibnamefont {Mackenzie}},\ }\bibfield  {title} {\bibinfo {title} {Controlled introduction of defects to delafossite metals by electron irradiation},\ }\href {https://doi.org/10.1103/PhysRevX.10.021018} {\bibfield  {journal} {\bibinfo
  {journal} {Phys. Rev. X}\ }\textbf {\bibinfo {volume} {10}},\ \bibinfo {pages} {021018} (\bibinfo {year} {2020})}\BibitemShut {NoStop}%
\bibitem [{\citenamefont {Legris}\ \emph {et~al.}(1993)\citenamefont {Legris}, \citenamefont {Rullier-Albenque}, \citenamefont {Radeva},\ and\ \citenamefont {Lejay}}]{legris1993}%
  \BibitemOpen
  \bibfield  {author} {\bibinfo {author} {\bibfnamefont {A.}~\bibnamefont {Legris}}, \bibinfo {author} {\bibfnamefont {F.}~\bibnamefont {Rullier-Albenque}}, \bibinfo {author} {\bibfnamefont {E.}~\bibnamefont {Radeva}},\ and\ \bibinfo {author} {\bibfnamefont {P.}~\bibnamefont {Lejay}},\ }\bibfield  {title} {\bibinfo {title} {{Effects of electron irradiation on {YBa}$_2${Cu}$_3${O}$_{7-\delta}$ superconductor}},\ }\href {https://doi.org/10.1051/jp1:1993203} {\bibfield  {journal} {\bibinfo  {journal} {{Journal de Physique I}}\ }\textbf {\bibinfo {volume} {3}},\ \bibinfo {pages} {1605} (\bibinfo {year} {1993})}\BibitemShut {NoStop}%
\bibitem [{\citenamefont {Rullier-Albenque}\ \emph {et~al.}(2003)\citenamefont {Rullier-Albenque}, \citenamefont {Alloul},\ and\ \citenamefont {Tourbot}}]{Rullier2003}%
  \BibitemOpen
  \bibfield  {author} {\bibinfo {author} {\bibfnamefont {F.}~\bibnamefont {Rullier-Albenque}}, \bibinfo {author} {\bibfnamefont {H.}~\bibnamefont {Alloul}},\ and\ \bibinfo {author} {\bibfnamefont {R.}~\bibnamefont {Tourbot}},\ }\bibfield  {title} {\bibinfo {title} {{Influence of Pair Breaking and Phase Fluctuations on Disordered High ${T}_{c}$ Cuprate Superconductors}},\ }\href {https://doi.org/10.1103/PhysRevLett.91.047001} {\bibfield  {journal} {\bibinfo  {journal} {Phys. Rev. Lett.}\ }\textbf {\bibinfo {volume} {91}},\ \bibinfo {pages} {047001} (\bibinfo {year} {2003})}\BibitemShut {NoStop}%
\bibitem [{\citenamefont {Openov}(2005)}]{Openov2005}%
  \BibitemOpen
  \bibfield  {author} {\bibinfo {author} {\bibfnamefont {L.~A.}\ \bibnamefont {Openov}},\ }\bibfield  {title} {\bibinfo {title} {{Irradiation-induced suppression of the critical temperature in high-${T}_{c}$ superconductors: Pair breaking versus phase fluctuations}},\ }\href {https://doi.org/10.1134/1.1881733} {\bibfield  {journal} {\bibinfo  {journal} {Journal of Experimental and Theoretical Physics Letters}\ }\textbf {\bibinfo {volume} {81}},\ \bibinfo {pages} {39} (\bibinfo {year} {2005})}\BibitemShut {NoStop}%
\bibitem [{\citenamefont {Ruf}\ \emph {et~al.}(2024)\citenamefont {Ruf}, \citenamefont {Noad}, \citenamefont {Grasset}, \citenamefont {Miao}, \citenamefont {Zhakina}, \citenamefont {McGuinness}, \citenamefont {Nair}, \citenamefont {Schreiber}, \citenamefont {Kikugawa}, \citenamefont {Sokolov}, \citenamefont {Konczykowski}, \citenamefont {Schlom}, \citenamefont {Shen},\ and\ \citenamefont {Mackenzie}}]{ruf_noad2024_SRO}%
  \BibitemOpen
  \bibfield  {author} {\bibinfo {author} {\bibfnamefont {J.~P.}\ \bibnamefont {Ruf}}, \bibinfo {author} {\bibfnamefont {H.~M.~L.}\ \bibnamefont {Noad}}, \bibinfo {author} {\bibfnamefont {R.}~\bibnamefont {Grasset}}, \bibinfo {author} {\bibfnamefont {L.}~\bibnamefont {Miao}}, \bibinfo {author} {\bibfnamefont {E.}~\bibnamefont {Zhakina}}, \bibinfo {author} {\bibfnamefont {P.~H.}\ \bibnamefont {McGuinness}}, \bibinfo {author} {\bibfnamefont {H.~P.}\ \bibnamefont {Nair}}, \bibinfo {author} {\bibfnamefont {N.~J.}\ \bibnamefont {Schreiber}}, \bibinfo {author} {\bibfnamefont {N.}~\bibnamefont {Kikugawa}}, \bibinfo {author} {\bibfnamefont {D.}~\bibnamefont {Sokolov}}, \bibinfo {author} {\bibfnamefont {M.}~\bibnamefont {Konczykowski}}, \bibinfo {author} {\bibfnamefont {D.~G.}\ \bibnamefont {Schlom}}, \bibinfo {author} {\bibfnamefont {K.~M.}\ \bibnamefont {Shen}},\ and\ \bibinfo {author} {\bibfnamefont {A.~P.}\ \bibnamefont {Mackenzie}},\ }\bibfield  {title} {\bibinfo {title} {Controllable suppression of the
  unconventional superconductivity in bulk and thin-film \ce{Sr2RuO4} via high-energy electron irradiation},\ }\href {https://doi.org/10.1103/PhysRevResearch.6.033178} {\bibfield  {journal} {\bibinfo  {journal} {Phys. Rev. Res.}\ }\textbf {\bibinfo {volume} {6}},\ \bibinfo {pages} {033178} (\bibinfo {year} {2024})}\BibitemShut {NoStop}%
\bibitem [{\citenamefont {Analytis}\ \emph {et~al.}(2006)\citenamefont {Analytis}, \citenamefont {Ardavan}, \citenamefont {Blundell}, \citenamefont {Owen}, \citenamefont {Garman}, \citenamefont {Jeynes},\ and\ \citenamefont {Powell}}]{analytis2006effect}%
  \BibitemOpen
  \bibfield  {author} {\bibinfo {author} {\bibfnamefont {J.~G.}\ \bibnamefont {Analytis}}, \bibinfo {author} {\bibfnamefont {A.}~\bibnamefont {Ardavan}}, \bibinfo {author} {\bibfnamefont {S.~J.}\ \bibnamefont {Blundell}}, \bibinfo {author} {\bibfnamefont {R.~L.}\ \bibnamefont {Owen}}, \bibinfo {author} {\bibfnamefont {E.~F.}\ \bibnamefont {Garman}}, \bibinfo {author} {\bibfnamefont {C.}~\bibnamefont {Jeynes}},\ and\ \bibinfo {author} {\bibfnamefont {B.~J.}\ \bibnamefont {Powell}},\ }\bibfield  {title} {\bibinfo {title} {Effect of irradiation-induced disorder on the conductivity and critical temperature of the organic superconductor $\kappa$-({BEDT-TTF})$_2${Cu}({SCN})$_2$},\ }\href {https://doi.org/10.1103/PhysRevLett.96.177002} {\bibfield  {journal} {\bibinfo  {journal} {Phys. Rev. Lett.}\ }\textbf {\bibinfo {volume} {96}},\ \bibinfo {pages} {177002} (\bibinfo {year} {2006})}\BibitemShut {NoStop}%
\bibitem [{\citenamefont {Prozorov}\ \emph {et~al.}(2014)\citenamefont {Prozorov}, \citenamefont {Ko{\'n}czykowski}, \citenamefont {Tanatar}, \citenamefont {Thaler}, \citenamefont {Bud’ko}, \citenamefont {Canfield}, \citenamefont {Mishra},\ and\ \citenamefont {Hirschfeld}}]{prozorov2014effect}%
  \BibitemOpen
  \bibfield  {author} {\bibinfo {author} {\bibfnamefont {R.}~\bibnamefont {Prozorov}}, \bibinfo {author} {\bibfnamefont {M.}~\bibnamefont {Ko{\'n}czykowski}}, \bibinfo {author} {\bibfnamefont {M.~A.}\ \bibnamefont {Tanatar}}, \bibinfo {author} {\bibfnamefont {A.}~\bibnamefont {Thaler}}, \bibinfo {author} {\bibfnamefont {S.~L.}\ \bibnamefont {Bud’ko}}, \bibinfo {author} {\bibfnamefont {P.~C.}\ \bibnamefont {Canfield}}, \bibinfo {author} {\bibfnamefont {V.}~\bibnamefont {Mishra}},\ and\ \bibinfo {author} {\bibfnamefont {P.}~\bibnamefont {Hirschfeld}},\ }\bibfield  {title} {\bibinfo {title} {Effect of electron irradiation on superconductivity in single crystals of {Ba(Fe$_{1-x}$Ru$_x$)$_2$As$_2$ ($x$=0.24)}},\ }\href {https://doi.org/10.1103/PhysRevX.4.041032} {\bibfield  {journal} {\bibinfo  {journal} {Physical Review X}\ }\textbf {\bibinfo {volume} {4}},\ \bibinfo {pages} {041032} (\bibinfo {year} {2014})}\BibitemShut {NoStop}%
\bibitem [{\citenamefont {Roppongi}\ \emph {et~al.}(2023)\citenamefont {Roppongi}, \citenamefont {Ishihara}, \citenamefont {Tanaka}, \citenamefont {Ogawa}, \citenamefont {Okada}, \citenamefont {Liu}, \citenamefont {Mukasa}, \citenamefont {Mizukami}, \citenamefont {Uwatoko}, \citenamefont {Grasset}, \citenamefont {Konczykowski}, \citenamefont {Ortiz}, \citenamefont {Wilson}, \citenamefont {Hashimoto},\ and\ \citenamefont {Shibauchi}}]{Roppongi2023}%
  \BibitemOpen
  \bibfield  {author} {\bibinfo {author} {\bibfnamefont {M.}~\bibnamefont {Roppongi}}, \bibinfo {author} {\bibfnamefont {K.}~\bibnamefont {Ishihara}}, \bibinfo {author} {\bibfnamefont {Y.}~\bibnamefont {Tanaka}}, \bibinfo {author} {\bibfnamefont {K.}~\bibnamefont {Ogawa}}, \bibinfo {author} {\bibfnamefont {K.}~\bibnamefont {Okada}}, \bibinfo {author} {\bibfnamefont {S.}~\bibnamefont {Liu}}, \bibinfo {author} {\bibfnamefont {K.}~\bibnamefont {Mukasa}}, \bibinfo {author} {\bibfnamefont {Y.}~\bibnamefont {Mizukami}}, \bibinfo {author} {\bibfnamefont {Y.}~\bibnamefont {Uwatoko}}, \bibinfo {author} {\bibfnamefont {R.}~\bibnamefont {Grasset}}, \bibinfo {author} {\bibfnamefont {M.}~\bibnamefont {Konczykowski}}, \bibinfo {author} {\bibfnamefont {B.~R.}\ \bibnamefont {Ortiz}}, \bibinfo {author} {\bibfnamefont {S.~D.}\ \bibnamefont {Wilson}}, \bibinfo {author} {\bibfnamefont {K.}~\bibnamefont {Hashimoto}},\ and\ \bibinfo {author} {\bibfnamefont {T.}~\bibnamefont {Shibauchi}},\ }\bibfield  {title} {\bibinfo {title}
  {{Bulk evidence of anisotropic $s$-wave pairing with no sign change in the kagome superconductor \ce{CsV3Sb5}}},\ }\href {https://doi.org/10.1038/s41467-023-36273-x} {\bibfield  {journal} {\bibinfo  {journal} {Nature Communications}\ }\textbf {\bibinfo {volume} {14}},\ \bibinfo {pages} {667} (\bibinfo {year} {2023})}\BibitemShut {NoStop}%
\bibitem [{\citenamefont {Lin}\ \emph {et~al.}(2015)\citenamefont {Lin}, \citenamefont {Rischau}, \citenamefont {van~der Beek}, \citenamefont {Fauqu\'e},\ and\ \citenamefont {Behnia}}]{Xiao2015_dopedSTO}%
  \BibitemOpen
  \bibfield  {author} {\bibinfo {author} {\bibfnamefont {X.}~\bibnamefont {Lin}}, \bibinfo {author} {\bibfnamefont {C.~W.}\ \bibnamefont {Rischau}}, \bibinfo {author} {\bibfnamefont {C.~J.}\ \bibnamefont {van~der Beek}}, \bibinfo {author} {\bibfnamefont {B.}~\bibnamefont {Fauqu\'e}},\ and\ \bibinfo {author} {\bibfnamefont {K.}~\bibnamefont {Behnia}},\ }\bibfield  {title} {\bibinfo {title} {{$s$-wave superconductivity in optimally doped {SrTi}$_{1-x}${Nb}$_x$O$_3$ unveiled by electron irradiation}},\ }\href {https://doi.org/10.1103/PhysRevB.92.174504} {\bibfield  {journal} {\bibinfo  {journal} {Phys. Rev. B}\ }\textbf {\bibinfo {volume} {92}},\ \bibinfo {pages} {174504} (\bibinfo {year} {2015})}\BibitemShut {NoStop}%
\bibitem [{\citenamefont {Ghimire}\ \emph {et~al.}(2024)\citenamefont {Ghimire}, \citenamefont {Joshi}, \citenamefont {Krenkel}, \citenamefont {Tanatar}, \citenamefont {Shi}, \citenamefont {Ko\ifmmode~\acute{n}\else \'{n}\fi{}czykowski}, \citenamefont {Grasset}, \citenamefont {Taufour}, \citenamefont {Orth}, \citenamefont {Scheurer},\ and\ \citenamefont {Prozorov}}]{Ghimire2024_LaNiGa2}%
  \BibitemOpen
  \bibfield  {author} {\bibinfo {author} {\bibfnamefont {S.}~\bibnamefont {Ghimire}}, \bibinfo {author} {\bibfnamefont {K.~R.}\ \bibnamefont {Joshi}}, \bibinfo {author} {\bibfnamefont {E.~H.}\ \bibnamefont {Krenkel}}, \bibinfo {author} {\bibfnamefont {M.~A.}\ \bibnamefont {Tanatar}}, \bibinfo {author} {\bibfnamefont {Y.}~\bibnamefont {Shi}}, \bibinfo {author} {\bibfnamefont {M.}~\bibnamefont {Ko\ifmmode~\acute{n}\else \'{n}\fi{}czykowski}}, \bibinfo {author} {\bibfnamefont {R.}~\bibnamefont {Grasset}}, \bibinfo {author} {\bibfnamefont {V.}~\bibnamefont {Taufour}}, \bibinfo {author} {\bibfnamefont {P.~P.}\ \bibnamefont {Orth}}, \bibinfo {author} {\bibfnamefont {M.~S.}\ \bibnamefont {Scheurer}},\ and\ \bibinfo {author} {\bibfnamefont {R.}~\bibnamefont {Prozorov}},\ }\bibfield  {title} {\bibinfo {title} {Electron irradiation reveals robust fully gapped superconductivity in {LaNiGa}$_{2}$},\ }\href {https://doi.org/10.1103/PhysRevB.109.024515} {\bibfield  {journal} {\bibinfo  {journal} {Phys. Rev. B}\ }\textbf
  {\bibinfo {volume} {109}},\ \bibinfo {pages} {024515} (\bibinfo {year} {2024})}\BibitemShut {NoStop}%
\bibitem [{\citenamefont {Lee}\ \emph {et~al.}(2020)\citenamefont {Lee}, \citenamefont {Goodge}, \citenamefont {Li}, \citenamefont {Osada}, \citenamefont {Wang}, \citenamefont {Cui}, \citenamefont {Kourkoutis},\ and\ \citenamefont {Hwang}}]{Lee2020_APL}%
  \BibitemOpen
  \bibfield  {author} {\bibinfo {author} {\bibfnamefont {K.}~\bibnamefont {Lee}}, \bibinfo {author} {\bibfnamefont {B.~H.}\ \bibnamefont {Goodge}}, \bibinfo {author} {\bibfnamefont {D.}~\bibnamefont {Li}}, \bibinfo {author} {\bibfnamefont {M.}~\bibnamefont {Osada}}, \bibinfo {author} {\bibfnamefont {B.~Y.}\ \bibnamefont {Wang}}, \bibinfo {author} {\bibfnamefont {Y.}~\bibnamefont {Cui}}, \bibinfo {author} {\bibfnamefont {L.~F.}\ \bibnamefont {Kourkoutis}},\ and\ \bibinfo {author} {\bibfnamefont {H.~Y.}\ \bibnamefont {Hwang}},\ }\bibfield  {title} {\bibinfo {title} {{Aspects of the synthesis of thin film superconducting infinite-layer nickelates}},\ }\href {https://doi.org/10.1063/5.0005103} {\bibfield  {journal} {\bibinfo  {journal} {APL Materials}\ }\textbf {\bibinfo {volume} {8}},\ \bibinfo {pages} {041107} (\bibinfo {year} {2020})}\BibitemShut {NoStop}%
\bibitem [{\citenamefont {Lee}\ \emph {et~al.}(2023)\citenamefont {Lee}, \citenamefont {Wang}, \citenamefont {Osada}, \citenamefont {Goodge}, \citenamefont {Wang}, \citenamefont {Lee}, \citenamefont {Harvey}, \citenamefont {Kim}, \citenamefont {Yu}, \citenamefont {Murthy}, \citenamefont {Raghu}, \citenamefont {Kourkoutis},\ and\ \citenamefont {Hwang}}]{Lee2023_normal_state}%
  \BibitemOpen
  \bibfield  {author} {\bibinfo {author} {\bibfnamefont {K.}~\bibnamefont {Lee}}, \bibinfo {author} {\bibfnamefont {B.~Y.}\ \bibnamefont {Wang}}, \bibinfo {author} {\bibfnamefont {M.}~\bibnamefont {Osada}}, \bibinfo {author} {\bibfnamefont {B.~H.}\ \bibnamefont {Goodge}}, \bibinfo {author} {\bibfnamefont {T.~C.}\ \bibnamefont {Wang}}, \bibinfo {author} {\bibfnamefont {Y.}~\bibnamefont {Lee}}, \bibinfo {author} {\bibfnamefont {S.}~\bibnamefont {Harvey}}, \bibinfo {author} {\bibfnamefont {W.~J.}\ \bibnamefont {Kim}}, \bibinfo {author} {\bibfnamefont {Y.}~\bibnamefont {Yu}}, \bibinfo {author} {\bibfnamefont {C.}~\bibnamefont {Murthy}}, \bibinfo {author} {\bibfnamefont {S.}~\bibnamefont {Raghu}}, \bibinfo {author} {\bibfnamefont {L.~F.}\ \bibnamefont {Kourkoutis}},\ and\ \bibinfo {author} {\bibfnamefont {H.~Y.}\ \bibnamefont {Hwang}},\ }\bibfield  {title} {\bibinfo {title} {Linear-in-temperature resistivity for optimally superconducting ({Nd, Sr)NiO$_2$}},\ }\href {https://doi.org/10.1038/s41586-023-06129-x}
  {\bibfield  {journal} {\bibinfo  {journal} {Nature}\ }\textbf {\bibinfo {volume} {619}},\ \bibinfo {pages} {288–292} (\bibinfo {year} {2023})}\BibitemShut {NoStop}%
\bibitem [{\citenamefont {Gonzalez}\ \emph {et~al.}(2024)\citenamefont {Gonzalez}, \citenamefont {Ievlev}, \citenamefont {Lee}, \citenamefont {Kim}, \citenamefont {Yu}, \citenamefont {Fowlie},\ and\ \citenamefont {Hwang}}]{Gonzalez2024}%
  \BibitemOpen
  \bibfield  {author} {\bibinfo {author} {\bibfnamefont {M.}~\bibnamefont {Gonzalez}}, \bibinfo {author} {\bibfnamefont {A.}~\bibnamefont {Ievlev}}, \bibinfo {author} {\bibfnamefont {K.}~\bibnamefont {Lee}}, \bibinfo {author} {\bibfnamefont {W.}~\bibnamefont {Kim}}, \bibinfo {author} {\bibfnamefont {Y.}~\bibnamefont {Yu}}, \bibinfo {author} {\bibfnamefont {J.}~\bibnamefont {Fowlie}},\ and\ \bibinfo {author} {\bibfnamefont {H.~Y.}\ \bibnamefont {Hwang}},\ }\bibfield  {title} {\bibinfo {title} {Absence of hydrogen insertion into highly crystalline superconducting infinite layer nickelates},\ }\href {https://doi.org/10.1103/PhysRevMaterials.8.084804} {\bibfield  {journal} {\bibinfo  {journal} {Phys. Rev. Mater.}\ }\textbf {\bibinfo {volume} {8}},\ \bibinfo {pages} {084804} (\bibinfo {year} {2024})}\BibitemShut {NoStop}%
\bibitem [{\citenamefont {Rullier-Albenque}\ \emph {et~al.}(2001)\citenamefont {Rullier-Albenque}, \citenamefont {Alloul},\ and\ \citenamefont {Tourbot}}]{Rullier2001}%
  \BibitemOpen
  \bibfield  {author} {\bibinfo {author} {\bibfnamefont {F.}~\bibnamefont {Rullier-Albenque}}, \bibinfo {author} {\bibfnamefont {H.}~\bibnamefont {Alloul}},\ and\ \bibinfo {author} {\bibfnamefont {R.}~\bibnamefont {Tourbot}},\ }\bibfield  {title} {\bibinfo {title} {{Disorder and Transport in Cuprates: Weak Localization and Magnetic Contributions}},\ }\href {https://doi.org/10.1103/PhysRevLett.87.157001} {\bibfield  {journal} {\bibinfo  {journal} {Phys. Rev. Lett.}\ }\textbf {\bibinfo {volume} {87}},\ \bibinfo {pages} {157001} (\bibinfo {year} {2001})}\BibitemShut {NoStop}%
\bibitem [{\citenamefont {Sun}\ \emph {et~al.}(1994)\citenamefont {Sun}, \citenamefont {Paulius}, \citenamefont {Gajewski}, \citenamefont {Maple},\ and\ \citenamefont {Dynes}}]{sun1994electron}%
  \BibitemOpen
  \bibfield  {author} {\bibinfo {author} {\bibfnamefont {A.~G.}\ \bibnamefont {Sun}}, \bibinfo {author} {\bibfnamefont {L.~M.}\ \bibnamefont {Paulius}}, \bibinfo {author} {\bibfnamefont {D.~A.}\ \bibnamefont {Gajewski}}, \bibinfo {author} {\bibfnamefont {M.~B.}\ \bibnamefont {Maple}},\ and\ \bibinfo {author} {\bibfnamefont {R.~C.}\ \bibnamefont {Dynes}},\ }\bibfield  {title} {\bibinfo {title} {Electron tunneling and transport in the high-{T}$_\textrm{c}$ superconductor {Y$_{1-x}$Pr$_x$Ba$_2$Cu$_3$O$_{7-\delta}$}},\ }\href {https://doi.org/10.1103/PhysRevB.50.3266} {\bibfield  {journal} {\bibinfo  {journal} {Phys. Rev. B}\ }\textbf {\bibinfo {volume} {50}},\ \bibinfo {pages} {3266} (\bibinfo {year} {1994})}\BibitemShut {NoStop}%
\bibitem [{\citenamefont {Bruin}\ \emph {et~al.}(2013)\citenamefont {Bruin}, \citenamefont {Sakai}, \citenamefont {Perry},\ and\ \citenamefont {Mackenzie}}]{Bruin_Mackenzie2013_Science}%
  \BibitemOpen
  \bibfield  {author} {\bibinfo {author} {\bibfnamefont {J.~A.~N.}\ \bibnamefont {Bruin}}, \bibinfo {author} {\bibfnamefont {H.}~\bibnamefont {Sakai}}, \bibinfo {author} {\bibfnamefont {R.~S.}\ \bibnamefont {Perry}},\ and\ \bibinfo {author} {\bibfnamefont {A.~P.}\ \bibnamefont {Mackenzie}},\ }\bibfield  {title} {\bibinfo {title} {{Similarity of Scattering Rates in Metals Showing $T$-Linear Resistivity}},\ }\href {https://doi.org/10.1126/science.1227612} {\bibfield  {journal} {\bibinfo  {journal} {Science}\ }\textbf {\bibinfo {volume} {339}},\ \bibinfo {pages} {804} (\bibinfo {year} {2013})}\BibitemShut {NoStop}%
\bibitem [{\citenamefont {Hartnoll}\ and\ \citenamefont {Mackenzie}(2022)}]{Hartnoll_Mackenzie2022_RMP}%
  \BibitemOpen
  \bibfield  {author} {\bibinfo {author} {\bibfnamefont {S.~A.}\ \bibnamefont {Hartnoll}}\ and\ \bibinfo {author} {\bibfnamefont {A.~P.}\ \bibnamefont {Mackenzie}},\ }\bibfield  {title} {\bibinfo {title} {{Colloquium: Planckian dissipation in metals}},\ }\href {https://doi.org/10.1103/RevModPhys.94.041002} {\bibfield  {journal} {\bibinfo  {journal} {Rev. Mod. Phys.}\ }\textbf {\bibinfo {volume} {94}},\ \bibinfo {pages} {041002} (\bibinfo {year} {2022})}\BibitemShut {NoStop}%
\bibitem [{\citenamefont {Varma}(2020)}]{Varma2020_T-Linear_RMP}%
  \BibitemOpen
  \bibfield  {author} {\bibinfo {author} {\bibfnamefont {C.~M.}\ \bibnamefont {Varma}},\ }\bibfield  {title} {\bibinfo {title} {{Colloquium: Linear in temperature resistivity and associated mysteries including high temperature superconductivity}},\ }\href {https://doi.org/10.1103/RevModPhys.92.031001} {\bibfield  {journal} {\bibinfo  {journal} {Rev. Mod. Phys.}\ }\textbf {\bibinfo {volume} {92}},\ \bibinfo {pages} {031001} (\bibinfo {year} {2020})}\BibitemShut {NoStop}%
\bibitem [{\citenamefont {Martin}\ \emph {et~al.}(1990)\citenamefont {Martin}, \citenamefont {Fiory}, \citenamefont {Fleming}, \citenamefont {Schneemeyer},\ and\ \citenamefont {Waszczak}}]{martin1990normal}%
  \BibitemOpen
  \bibfield  {author} {\bibinfo {author} {\bibfnamefont {S.}~\bibnamefont {Martin}}, \bibinfo {author} {\bibfnamefont {A.~T.}\ \bibnamefont {Fiory}}, \bibinfo {author} {\bibfnamefont {R.}~\bibnamefont {Fleming}}, \bibinfo {author} {\bibfnamefont {L.}~\bibnamefont {Schneemeyer}},\ and\ \bibinfo {author} {\bibfnamefont {J.~V.}\ \bibnamefont {Waszczak}},\ }\bibfield  {title} {\bibinfo {title} {Normal-state transport properties of {Bi}$_{2+x}${Sr}$_{2-y}${CuO}$_{6+\delta}$ crystals},\ }\href {https://doi.org/10.1103/PhysRevB.41.846} {\bibfield  {journal} {\bibinfo  {journal} {Physical Review B}\ }\textbf {\bibinfo {volume} {41}},\ \bibinfo {pages} {846} (\bibinfo {year} {1990})}\BibitemShut {NoStop}%
\bibitem [{\citenamefont {Doiron-Leyraud}\ \emph {et~al.}(2009)\citenamefont {Doiron-Leyraud}, \citenamefont {Auban-Senzier}, \citenamefont {Ren{\'e}~de Cotret}, \citenamefont {Bourbonnais}, \citenamefont {J{\'e}rome}, \citenamefont {Bechgaard},\ and\ \citenamefont {Taillefer}}]{doiron2009correlation}%
  \BibitemOpen
  \bibfield  {author} {\bibinfo {author} {\bibfnamefont {N.}~\bibnamefont {Doiron-Leyraud}}, \bibinfo {author} {\bibfnamefont {P.}~\bibnamefont {Auban-Senzier}}, \bibinfo {author} {\bibfnamefont {S.}~\bibnamefont {Ren{\'e}~de Cotret}}, \bibinfo {author} {\bibfnamefont {C.}~\bibnamefont {Bourbonnais}}, \bibinfo {author} {\bibfnamefont {D.}~\bibnamefont {J{\'e}rome}}, \bibinfo {author} {\bibfnamefont {K.}~\bibnamefont {Bechgaard}},\ and\ \bibinfo {author} {\bibfnamefont {L.}~\bibnamefont {Taillefer}},\ }\bibfield  {title} {\bibinfo {title} {Correlation between linear resistivity and ${T}_\textrm{c}$ in the {B}echgaard salts and the pnictide superconductor {Ba}({Fe}$_{1-x}${Co}$_x$)$_2${As}$_2$},\ }\href {https://doi.org/10.1103/PhysRevB.80.214531} {\bibfield  {journal} {\bibinfo  {journal} {Physical Review B—Condensed Matter and Materials Physics}\ }\textbf {\bibinfo {volume} {80}},\ \bibinfo {pages} {214531} (\bibinfo {year} {2009})}\BibitemShut {NoStop}%
\bibitem [{\citenamefont {L\"ohneysen}\ \emph {et~al.}(1994)\citenamefont {L\"ohneysen}, \citenamefont {Pietrus}, \citenamefont {Portisch}, \citenamefont {Schlager}, \citenamefont {Schr\"oder}, \citenamefont {Sieck},\ and\ \citenamefont {Trappmann}}]{Lohneysen1994}%
  \BibitemOpen
  \bibfield  {author} {\bibinfo {author} {\bibfnamefont {H.~v.}\ \bibnamefont {L\"ohneysen}}, \bibinfo {author} {\bibfnamefont {T.}~\bibnamefont {Pietrus}}, \bibinfo {author} {\bibfnamefont {G.}~\bibnamefont {Portisch}}, \bibinfo {author} {\bibfnamefont {H.~G.}\ \bibnamefont {Schlager}}, \bibinfo {author} {\bibfnamefont {A.}~\bibnamefont {Schr\"oder}}, \bibinfo {author} {\bibfnamefont {M.}~\bibnamefont {Sieck}},\ and\ \bibinfo {author} {\bibfnamefont {T.}~\bibnamefont {Trappmann}},\ }\bibfield  {title} {\bibinfo {title} {{Non-Fermi-liquid behavior in a heavy-fermion alloy at a magnetic instability}},\ }\href {https://doi.org/10.1103/PhysRevLett.72.3262} {\bibfield  {journal} {\bibinfo  {journal} {Phys. Rev. Lett.}\ }\textbf {\bibinfo {volume} {72}},\ \bibinfo {pages} {3262} (\bibinfo {year} {1994})}\BibitemShut {NoStop}%
\bibitem [{\citenamefont {Anderson}\ and\ \citenamefont {Zou}(1988)}]{Anderson1988_VBS_Transport}%
  \BibitemOpen
  \bibfield  {author} {\bibinfo {author} {\bibfnamefont {P.~W.}\ \bibnamefont {Anderson}}\ and\ \bibinfo {author} {\bibfnamefont {Z.}~\bibnamefont {Zou}},\ }\bibfield  {title} {\bibinfo {title} {{``Normal" Tunneling and ``Normal" Transport: Diagnostics for the Resonating-Valence-Bond State}},\ }\href {https://doi.org/10.1103/PhysRevLett.60.132} {\bibfield  {journal} {\bibinfo  {journal} {Phys. Rev. Lett.}\ }\textbf {\bibinfo {volume} {60}},\ \bibinfo {pages} {132} (\bibinfo {year} {1988})}\BibitemShut {NoStop}%
\bibitem [{\citenamefont {Orenstein}\ \emph {et~al.}(1990)\citenamefont {Orenstein}, \citenamefont {Thomas}, \citenamefont {Millis}, \citenamefont {Cooper}, \citenamefont {Rapkine}, \citenamefont {Timusk}, \citenamefont {Schneemeyer},\ and\ \citenamefont {Waszczak}}]{Orenstein1990}%
  \BibitemOpen
  \bibfield  {author} {\bibinfo {author} {\bibfnamefont {J.}~\bibnamefont {Orenstein}}, \bibinfo {author} {\bibfnamefont {G.~A.}\ \bibnamefont {Thomas}}, \bibinfo {author} {\bibfnamefont {A.~J.}\ \bibnamefont {Millis}}, \bibinfo {author} {\bibfnamefont {S.~L.}\ \bibnamefont {Cooper}}, \bibinfo {author} {\bibfnamefont {D.~H.}\ \bibnamefont {Rapkine}}, \bibinfo {author} {\bibfnamefont {T.}~\bibnamefont {Timusk}}, \bibinfo {author} {\bibfnamefont {L.~F.}\ \bibnamefont {Schneemeyer}},\ and\ \bibinfo {author} {\bibfnamefont {J.~V.}\ \bibnamefont {Waszczak}},\ }\bibfield  {title} {\bibinfo {title} {Frequency-and temperature-dependent conductivity in {YBa}$_2${Cu}$_3${O}$_{6+x}$ crystals},\ }\href {https://doi.org/10.1103/PhysRevB.42.6342} {\bibfield  {journal} {\bibinfo  {journal} {Phys. Rev. B}\ }\textbf {\bibinfo {volume} {42}},\ \bibinfo {pages} {6342} (\bibinfo {year} {1990})}\BibitemShut {NoStop}%
\bibitem [{\citenamefont {Marel}\ \emph {et~al.}(2003)\citenamefont {Marel}, \citenamefont {Molegraaf}, \citenamefont {Zaanen}, \citenamefont {Nussinov}, \citenamefont {Carbone}, \citenamefont {Damascelli}, \citenamefont {Eisaki}, \citenamefont {Greven}, \citenamefont {Kes},\ and\ \citenamefont {Li}}]{Marel2003}%
  \BibitemOpen
  \bibfield  {author} {\bibinfo {author} {\bibfnamefont {D.~v.~d.}\ \bibnamefont {Marel}}, \bibinfo {author} {\bibfnamefont {H.~J.~A.}\ \bibnamefont {Molegraaf}}, \bibinfo {author} {\bibfnamefont {J.}~\bibnamefont {Zaanen}}, \bibinfo {author} {\bibfnamefont {Z.}~\bibnamefont {Nussinov}}, \bibinfo {author} {\bibfnamefont {F.}~\bibnamefont {Carbone}}, \bibinfo {author} {\bibfnamefont {A.}~\bibnamefont {Damascelli}}, \bibinfo {author} {\bibfnamefont {H.}~\bibnamefont {Eisaki}}, \bibinfo {author} {\bibfnamefont {M.}~\bibnamefont {Greven}}, \bibinfo {author} {\bibfnamefont {P.~H.}\ \bibnamefont {Kes}},\ and\ \bibinfo {author} {\bibfnamefont {M.}~\bibnamefont {Li}},\ }\bibfield  {title} {\bibinfo {title} {{Quantum critical behaviour in a high-$T_c$ superconductor}},\ }\href {https://doi.org/10.1038/nature01978} {\bibfield  {journal} {\bibinfo  {journal} {Nature}\ }\textbf {\bibinfo {volume} {425}},\ \bibinfo {pages} {271} (\bibinfo {year} {2003})}\BibitemShut {NoStop}%
\bibitem [{\citenamefont {Zaanen}(2004)}]{Zaanen2004}%
  \BibitemOpen
  \bibfield  {author} {\bibinfo {author} {\bibfnamefont {J.}~\bibnamefont {Zaanen}},\ }\bibfield  {title} {\bibinfo {title} {Why the temperature is high},\ }\href {https://doi.org/10.1038/430512a} {\bibfield  {journal} {\bibinfo  {journal} {Nature}\ }\textbf {\bibinfo {volume} {430}},\ \bibinfo {pages} {512} (\bibinfo {year} {2004})}\BibitemShut {NoStop}%
\bibitem [{\citenamefont {Takagi}\ \emph {et~al.}(1992)\citenamefont {Takagi}, \citenamefont {Batlogg}, \citenamefont {Kao}, \citenamefont {Kwo}, \citenamefont {Cava}, \citenamefont {Krajewski},\ and\ \citenamefont {Peck}}]{Takagi1992_LSCO}%
  \BibitemOpen
  \bibfield  {author} {\bibinfo {author} {\bibfnamefont {H.}~\bibnamefont {Takagi}}, \bibinfo {author} {\bibfnamefont {B.}~\bibnamefont {Batlogg}}, \bibinfo {author} {\bibfnamefont {H.~L.}\ \bibnamefont {Kao}}, \bibinfo {author} {\bibfnamefont {J.}~\bibnamefont {Kwo}}, \bibinfo {author} {\bibfnamefont {R.~J.}\ \bibnamefont {Cava}}, \bibinfo {author} {\bibfnamefont {J.~J.}\ \bibnamefont {Krajewski}},\ and\ \bibinfo {author} {\bibfnamefont {W.~F.}\ \bibnamefont {Peck}},\ }\bibfield  {title} {\bibinfo {title} {{Systematic evolution of temperature-dependent resistivity in La$_{2-x}$Sr$_x$CuO$_4$}},\ }\href {https://doi.org/10.1103/PhysRevLett.69.2975} {\bibfield  {journal} {\bibinfo  {journal} {Phys. Rev. Lett.}\ }\textbf {\bibinfo {volume} {69}},\ \bibinfo {pages} {2975} (\bibinfo {year} {1992})}\BibitemShut {NoStop}%
\bibitem [{\citenamefont {Boebinger}\ \emph {et~al.}(1996)\citenamefont {Boebinger}, \citenamefont {Ando}, \citenamefont {Passner}, \citenamefont {Kimura}, \citenamefont {Okuya}, \citenamefont {Shimoyama}, \citenamefont {Kishio}, \citenamefont {Tamasaku}, \citenamefont {Ichikawa},\ and\ \citenamefont {Uchida}}]{Boebinger1996_LSCO}%
  \BibitemOpen
  \bibfield  {author} {\bibinfo {author} {\bibfnamefont {G.~S.}\ \bibnamefont {Boebinger}}, \bibinfo {author} {\bibfnamefont {Y.}~\bibnamefont {Ando}}, \bibinfo {author} {\bibfnamefont {A.}~\bibnamefont {Passner}}, \bibinfo {author} {\bibfnamefont {T.}~\bibnamefont {Kimura}}, \bibinfo {author} {\bibfnamefont {M.}~\bibnamefont {Okuya}}, \bibinfo {author} {\bibfnamefont {J.}~\bibnamefont {Shimoyama}}, \bibinfo {author} {\bibfnamefont {K.}~\bibnamefont {Kishio}}, \bibinfo {author} {\bibfnamefont {K.}~\bibnamefont {Tamasaku}}, \bibinfo {author} {\bibfnamefont {N.}~\bibnamefont {Ichikawa}},\ and\ \bibinfo {author} {\bibfnamefont {S.}~\bibnamefont {Uchida}},\ }\bibfield  {title} {\bibinfo {title} {{Insulator-to-Metal Crossover in the Normal State of La$_{2-x}$Sr$_x$CuO$_4$ Near Optimum Doping}},\ }\href {https://doi.org/10.1103/PhysRevLett.77.5417} {\bibfield  {journal} {\bibinfo  {journal} {Phys. Rev. Lett.}\ }\textbf {\bibinfo {volume} {77}},\ \bibinfo {pages} {5417} (\bibinfo {year} {1996})}\BibitemShut
  {NoStop}%
\bibitem [{\citenamefont {Cooper}\ \emph {et~al.}(2009)\citenamefont {Cooper}, \citenamefont {Wang}, \citenamefont {Vignolle}, \citenamefont {Lipscombe}, \citenamefont {Hayden}, \citenamefont {Tanabe}, \citenamefont {Adachi}, \citenamefont {Koike}, \citenamefont {Nohara}, \citenamefont {Takagi}, \citenamefont {Proust},\ and\ \citenamefont {Hussey}}]{Cooper2009_LSCO}%
  \BibitemOpen
  \bibfield  {author} {\bibinfo {author} {\bibfnamefont {R.~A.}\ \bibnamefont {Cooper}}, \bibinfo {author} {\bibfnamefont {Y.}~\bibnamefont {Wang}}, \bibinfo {author} {\bibfnamefont {B.}~\bibnamefont {Vignolle}}, \bibinfo {author} {\bibfnamefont {O.~J.}\ \bibnamefont {Lipscombe}}, \bibinfo {author} {\bibfnamefont {S.~M.}\ \bibnamefont {Hayden}}, \bibinfo {author} {\bibfnamefont {Y.}~\bibnamefont {Tanabe}}, \bibinfo {author} {\bibfnamefont {T.}~\bibnamefont {Adachi}}, \bibinfo {author} {\bibfnamefont {Y.}~\bibnamefont {Koike}}, \bibinfo {author} {\bibfnamefont {M.}~\bibnamefont {Nohara}}, \bibinfo {author} {\bibfnamefont {H.}~\bibnamefont {Takagi}}, \bibinfo {author} {\bibfnamefont {C.}~\bibnamefont {Proust}},\ and\ \bibinfo {author} {\bibfnamefont {N.~E.}\ \bibnamefont {Hussey}},\ }\bibfield  {title} {\bibinfo {title} {{Anomalous Criticality in the Electrical Resistivity of La$_{2-x}$Sr$_x$CuO$_4$}},\ }\href {https://doi.org/10.1126/science.1165015} {\bibfield  {journal} {\bibinfo  {journal} {Science}\
  }\textbf {\bibinfo {volume} {323}},\ \bibinfo {pages} {603} (\bibinfo {year} {2009})}\BibitemShut {NoStop}%
\bibitem [{\citenamefont {Mizukami}\ \emph {et~al.}(2014)\citenamefont {Mizukami}, \citenamefont {Konczykowski}, \citenamefont {Kawamoto}, \citenamefont {Kurata}, \citenamefont {Kasahara}, \citenamefont {Hashimoto}, \citenamefont {Mishra}, \citenamefont {Kreisel}, \citenamefont {Wang}, \citenamefont {Hirschfeld}, \citenamefont {Matsuda},\ and\ \citenamefont {Shibauchi}}]{Mizukami2014_iron_pnictide}%
  \BibitemOpen
  \bibfield  {author} {\bibinfo {author} {\bibfnamefont {Y.}~\bibnamefont {Mizukami}}, \bibinfo {author} {\bibfnamefont {M.}~\bibnamefont {Konczykowski}}, \bibinfo {author} {\bibfnamefont {Y.}~\bibnamefont {Kawamoto}}, \bibinfo {author} {\bibfnamefont {S.}~\bibnamefont {Kurata}}, \bibinfo {author} {\bibfnamefont {S.}~\bibnamefont {Kasahara}}, \bibinfo {author} {\bibfnamefont {K.}~\bibnamefont {Hashimoto}}, \bibinfo {author} {\bibfnamefont {V.}~\bibnamefont {Mishra}}, \bibinfo {author} {\bibfnamefont {A.}~\bibnamefont {Kreisel}}, \bibinfo {author} {\bibfnamefont {Y.}~\bibnamefont {Wang}}, \bibinfo {author} {\bibfnamefont {P.~J.}\ \bibnamefont {Hirschfeld}}, \bibinfo {author} {\bibfnamefont {Y.}~\bibnamefont {Matsuda}},\ and\ \bibinfo {author} {\bibfnamefont {T.}~\bibnamefont {Shibauchi}},\ }\bibfield  {title} {\bibinfo {title} {Disorder-induced topological change of the superconducting gap structure in iron pnictides},\ }\href {https://doi.org/10.1038/ncomms6657} {\bibfield  {journal} {\bibinfo  {journal}
  {Nature Communications}\ }\textbf {\bibinfo {volume} {5}},\ \bibinfo {pages} {5657} (\bibinfo {year} {2014})}\BibitemShut {NoStop}%
\bibitem [{\citenamefont {Larkin}(1965)}]{Larkin1965}%
  \BibitemOpen
  \bibfield  {author} {\bibinfo {author} {\bibfnamefont {A.~I.}\ \bibnamefont {Larkin}},\ }\bibfield  {title} {\bibinfo {title} {Effect of inhomogeneities on the structure of the mixed state of superconductors},\ }\href@noop {} {\bibfield  {journal} {\bibinfo  {journal} {Zh. Eksp. Teor. Fiz. Pis’ma Red.}\ }\textbf {\bibinfo {volume} {2}},\ \bibinfo {pages} {205} (\bibinfo {year} {1965})},\ \bibinfo {note} {[JETP Lett. \textbf{2}, 130 (1965)]}\BibitemShut {NoStop}%
\bibitem [{\citenamefont {Millis}\ \emph {et~al.}(1988)\citenamefont {Millis}, \citenamefont {Sachdev},\ and\ \citenamefont {Varma}}]{Millis1988}%
  \BibitemOpen
  \bibfield  {author} {\bibinfo {author} {\bibfnamefont {A.~J.}\ \bibnamefont {Millis}}, \bibinfo {author} {\bibfnamefont {S.}~\bibnamefont {Sachdev}},\ and\ \bibinfo {author} {\bibfnamefont {C.~M.}\ \bibnamefont {Varma}},\ }\bibfield  {title} {\bibinfo {title} {Inelastic scattering and pair breaking in anisotropic and isotropic superconductors},\ }\href {https://doi.org/10.1103/PhysRevB.37.4975} {\bibfield  {journal} {\bibinfo  {journal} {Phys. Rev. B}\ }\textbf {\bibinfo {volume} {37}},\ \bibinfo {pages} {4975} (\bibinfo {year} {1988})}\BibitemShut {NoStop}%
\bibitem [{\citenamefont {Anderson}(1959)}]{Anderson1959}%
  \BibitemOpen
  \bibfield  {author} {\bibinfo {author} {\bibfnamefont {P.}~\bibnamefont {Anderson}},\ }\bibfield  {title} {\bibinfo {title} {Theory of dirty superconductors},\ }\href {https://doi.org/https://doi.org/10.1016/0022-3697(59)90036-8} {\bibfield  {journal} {\bibinfo  {journal} {Journal of Physics and Chemistry of Solids}\ }\textbf {\bibinfo {volume} {11}},\ \bibinfo {pages} {26} (\bibinfo {year} {1959})}\BibitemShut {NoStop}%
\bibitem [{\citenamefont {Cavanagh}\ and\ \citenamefont {Brydon}(2020)}]{cavanagh2020_robust_s_wave}%
  \BibitemOpen
  \bibfield  {author} {\bibinfo {author} {\bibfnamefont {D.~C.}\ \bibnamefont {Cavanagh}}\ and\ \bibinfo {author} {\bibfnamefont {P.~M.~R.}\ \bibnamefont {Brydon}},\ }\bibfield  {title} {\bibinfo {title} {Robustness of unconventional $s$-wave superconducting states against disorder},\ }\href {https://doi.org/10.1103/PhysRevB.101.054509} {\bibfield  {journal} {\bibinfo  {journal} {Phys. Rev. B}\ }\textbf {\bibinfo {volume} {101}},\ \bibinfo {pages} {054509} (\bibinfo {year} {2020})}\BibitemShut {NoStop}%
\bibitem [{Note1()}]{Note1}%
  \BibitemOpen
  \bibinfo {note} {\textcolor {darkgreen}{ More explicitly, we focus on the second term of the first digamma function argument in Eq. 1 and start by substituting $\alpha = \protect \hbar / 2\tau k_B T_\protect \textrm {c0}$ to rewrite as \begin {equation} \protect \frac {\alpha T_\protect \textrm {c0} }{2 \pi T_\protect \textrm {c}} = \protect \frac {\protect \hbar }{4 \pi k_B T_\protect \textrm {c} \tau }. \label {eq:sub1} \end {equation} Further substituting the Drude expression for scattering time $1/\tau = (\omega ^2_{pl} / 4\pi ) \rho _0$ into the same term, we get \begin {equation} \protect \frac {\rho ^G_0}{T_\protect \textrm {c}}\protect \frac {\protect \hbar \omega _{pl}^2}{4 \pi k_B *4 \pi } \label {eq:sub2} \end {equation} where $\rho ^G_0$ denotes the resistivity in Gaussian CGS units (dimensions of time). For comparison to our experimental measurements, we convert into SI units by \begin {equation} \rho ^G_0 = \rho ^I_0 * 4 \pi \varepsilon _0 \label {eq:CGS} \end {equation} where $\rho ^I_0$ is
  the resistivity in SI units (dimensions of Ohm meters). Combining Eq. \ref {eq:sub2} with Eq. \ref {eq:CGS} and taking the assumption $\rho _0 \rightarrow \Delta \rho _0$ as discussed in the text, we obtain a simplified digamma argument term, $C \Delta \rho ^I_0 / T_\protect \textrm {c}$, where \begin {equation} C = \protect \frac {\varepsilon _0 \protect \hbar }{4 \pi k_B}\omega _{pl}^2 \label {eq:Cdef} \end {equation} in SI units, with dimensions of Kelvin per resistivity. All resistivities reported within the text are $\rho ^I$ }}\BibitemShut {NoStop}%
\bibitem [{Note2()}]{Note2}%
  \BibitemOpen
  \bibinfo {note} {\textcolor {darkgreen}{From optical measurements of Nd$_{0.8}$Sr$_{0.2}$NiO$_2$ films on SrTiO$_3$, Cervasio et al. \cite {cervasio2023optical} reported a room-temperature plasma frequency of $\sim 5500$ cm$^{-1}$ ($\protect \hbar \omega _{pl}\sim $0.68 eV), which would correspond to $C\sim 57$ K/m$\Omega \cdot $cm in our formulation. Encouragingly, this falls within the same order of magnitude as our fits to S1, though we note several challenges for more precise comparison, which include accounting for sample-to-sample variation with the reported transport data, temperature dependence, and low sensitivity of the fit dependence on $C$.}}\BibitemShut {Stop}%
\bibitem [{\citenamefont {Emery}\ and\ \citenamefont {Kivelson}(1995{\natexlab{a}})}]{EmeryKivelson_1995}%
  \BibitemOpen
  \bibfield  {author} {\bibinfo {author} {\bibfnamefont {V.~J.}\ \bibnamefont {Emery}}\ and\ \bibinfo {author} {\bibfnamefont {S.~A.}\ \bibnamefont {Kivelson}},\ }\bibfield  {title} {\bibinfo {title} {{Superconductivity in Bad Metals}},\ }\href {https://doi.org/10.1103/PhysRevLett.74.3253} {\bibfield  {journal} {\bibinfo  {journal} {Phys. Rev. Lett.}\ }\textbf {\bibinfo {volume} {74}},\ \bibinfo {pages} {3253} (\bibinfo {year} {1995}{\natexlab{a}})}\BibitemShut {NoStop}%
\bibitem [{\citenamefont {Emery}\ and\ \citenamefont {Kivelson}(1995{\natexlab{b}})}]{EmeryKivelson2_1995}%
  \BibitemOpen
  \bibfield  {author} {\bibinfo {author} {\bibfnamefont {V.~J.}\ \bibnamefont {Emery}}\ and\ \bibinfo {author} {\bibfnamefont {S.~A.}\ \bibnamefont {Kivelson}},\ }\bibfield  {title} {\bibinfo {title} {Importance of phase fluctuations in superconductors with small superfluid density},\ }\href {https://doi.org/10.1038/374434a0} {\bibfield  {journal} {\bibinfo  {journal} {Nature}\ }\textbf {\bibinfo {volume} {374}},\ \bibinfo {pages} {434} (\bibinfo {year} {1995}{\natexlab{b}})}\BibitemShut {NoStop}%
\bibitem [{Note3()}]{Note3}%
  \BibitemOpen
  \bibinfo {note} {In an ideal unconventional superconductor with only point defects, the critical mean free path $\lambda ^{\protect \textrm {crit}}$ for the appearance of superconductivity provides an estimate of the superconducting coherence length $\xi $. Notwithstanding the concerns about sample-to-sample variation of $\Delta \rho _0$ for a given radiation dose (Fig. \ref {fig2}b), one reasonable hypothesis would be to consider the lowest $\Delta \rho _0$ values (Fig. \ref {fig_ag}a) as the closest to the intrinsic values that would result from purely point defect scattering. It is then possible to estimate $\lambda ^{\protect \textrm {crit}}$ from $\Delta \rho _0^{\protect \textrm {crit}}$ $\protect \cong $ 60 {\textmu }$\Omega \cdot $cm \textcolor {darkgreen}{in sample S1}. In a sweeping approximation, we use the 2D expression $\lambda \protect \cong (1/\rho )(hd/(e^2 k_F)$ , where $d$ is the interlayer spacing and $k_F$ the Fermi wavevector, taking $k_F\sim $ 0.45 Å$^{-1}$ from recent photoemission
  experiments \cite {ding2024cuprate,Sun2025_LSNO_ARPES}, considering only the large quasi-two dimensional Fermi surface and neglecting the electron pockets near the zone corners. The resulting estimate of $\lambda ^\protect \textrm {crit}$ $\protect \cong $ $\xi $ $\protect \cong $ 33 Å is therefore only a rough one, but yields a value for $\xi $ in reasonable agreement with estimates from other measurements \cite {li_superconductivity_2019,wang2021isotropic}.}\BibitemShut {Stop}%
\bibitem [{\citenamefont {Franz}\ \emph {et~al.}(1997)\citenamefont {Franz}, \citenamefont {Kallin}, \citenamefont {Berlinsky},\ and\ \citenamefont {Salkola}}]{Franz1997_PRB}%
  \BibitemOpen
  \bibfield  {author} {\bibinfo {author} {\bibfnamefont {M.}~\bibnamefont {Franz}}, \bibinfo {author} {\bibfnamefont {C.}~\bibnamefont {Kallin}}, \bibinfo {author} {\bibfnamefont {A.~J.}\ \bibnamefont {Berlinsky}},\ and\ \bibinfo {author} {\bibfnamefont {M.~I.}\ \bibnamefont {Salkola}},\ }\bibfield  {title} {\bibinfo {title} {{Critical temperature and superfluid density suppression in disordered high-${T}_{c}$ cuprate superconductors}},\ }\href {https://doi.org/10.1103/PhysRevB.56.7882} {\bibfield  {journal} {\bibinfo  {journal} {Phys. Rev. B}\ }\textbf {\bibinfo {volume} {56}},\ \bibinfo {pages} {7882} (\bibinfo {year} {1997})}\BibitemShut {NoStop}%
\bibitem [{\citenamefont {Blinkin}\ \emph {et~al.}(2006)\citenamefont {Blinkin}, \citenamefont {Derevyanko}, \citenamefont {Dovbnya}, \citenamefont {Sukhareva}, \citenamefont {Finkel'},\ and\ \citenamefont {Shlyakhov}}]{Blinkin2006_MgB2_s-wave}%
  \BibitemOpen
  \bibfield  {author} {\bibinfo {author} {\bibfnamefont {A.~A.}\ \bibnamefont {Blinkin}}, \bibinfo {author} {\bibfnamefont {V.~V.}\ \bibnamefont {Derevyanko}}, \bibinfo {author} {\bibfnamefont {A.~N.}\ \bibnamefont {Dovbnya}}, \bibinfo {author} {\bibfnamefont {T.~V.}\ \bibnamefont {Sukhareva}}, \bibinfo {author} {\bibfnamefont {V.~A.}\ \bibnamefont {Finkel'}},\ and\ \bibinfo {author} {\bibfnamefont {I.~N.}\ \bibnamefont {Shlyakhov}},\ }\bibfield  {title} {\bibinfo {title} {Effect of electron irradiation on the structure and properties of the {MgB}$_2$ superconductor},\ }\href {https://doi.org/10.1134/S1063783406110011} {\bibfield  {journal} {\bibinfo  {journal} {Physics of the Solid State}\ }\textbf {\bibinfo {volume} {48}},\ \bibinfo {pages} {2037} (\bibinfo {year} {2006})}\BibitemShut {NoStop}%
\bibitem [{\citenamefont {Osada}\ \emph {et~al.}(2020)\citenamefont {Osada}, \citenamefont {Wang}, \citenamefont {Goodge}, \citenamefont {Lee}, \citenamefont {Yoon}, \citenamefont {Sakuma}, \citenamefont {Li}, \citenamefont {Miura}, \citenamefont {Kourkoutis},\ and\ \citenamefont {Hwang}}]{osada2020superconducting}%
  \BibitemOpen
  \bibfield  {author} {\bibinfo {author} {\bibfnamefont {M.}~\bibnamefont {Osada}}, \bibinfo {author} {\bibfnamefont {B.~Y.}\ \bibnamefont {Wang}}, \bibinfo {author} {\bibfnamefont {B.~H.}\ \bibnamefont {Goodge}}, \bibinfo {author} {\bibfnamefont {K.}~\bibnamefont {Lee}}, \bibinfo {author} {\bibfnamefont {H.}~\bibnamefont {Yoon}}, \bibinfo {author} {\bibfnamefont {K.}~\bibnamefont {Sakuma}}, \bibinfo {author} {\bibfnamefont {D.}~\bibnamefont {Li}}, \bibinfo {author} {\bibfnamefont {M.}~\bibnamefont {Miura}}, \bibinfo {author} {\bibfnamefont {L.~F.}\ \bibnamefont {Kourkoutis}},\ and\ \bibinfo {author} {\bibfnamefont {H.~Y.}\ \bibnamefont {Hwang}},\ }\bibfield  {title} {\bibinfo {title} {{A Superconducting Praseodymium Nickelate with Infinite Layer Structure}},\ }\href {https://doi.org/10.1021/acs.nanolett.0c01392} {\bibfield  {journal} {\bibinfo  {journal} {Nano Letters}\ }\textbf {\bibinfo {volume} {20}},\ \bibinfo {pages} {5735} (\bibinfo {year} {2020})}\BibitemShut {NoStop}%
\bibitem [{\citenamefont {Osada}\ \emph {et~al.}(2021)\citenamefont {Osada}, \citenamefont {Wang}, \citenamefont {Goodge}, \citenamefont {Harvey}, \citenamefont {Lee}, \citenamefont {Li}, \citenamefont {Kourkoutis},\ and\ \citenamefont {Hwang}}]{osada2021nickelate}%
  \BibitemOpen
  \bibfield  {author} {\bibinfo {author} {\bibfnamefont {M.}~\bibnamefont {Osada}}, \bibinfo {author} {\bibfnamefont {B.~Y.}\ \bibnamefont {Wang}}, \bibinfo {author} {\bibfnamefont {B.~H.}\ \bibnamefont {Goodge}}, \bibinfo {author} {\bibfnamefont {S.~P.}\ \bibnamefont {Harvey}}, \bibinfo {author} {\bibfnamefont {K.}~\bibnamefont {Lee}}, \bibinfo {author} {\bibfnamefont {D.}~\bibnamefont {Li}}, \bibinfo {author} {\bibfnamefont {L.~F.}\ \bibnamefont {Kourkoutis}},\ and\ \bibinfo {author} {\bibfnamefont {H.~Y.}\ \bibnamefont {Hwang}},\ }\bibfield  {title} {\bibinfo {title} {{Nickelate Superconductivity without Rare-Earth Magnetism: ({L}a,{S}r){N}i{O}$_2$}},\ }\href {https://onlinelibrary.wiley.com/doi/abs/10.1002/adma.202104083} {\bibfield  {journal} {\bibinfo  {journal} {Adv. Mater.}\ }\textbf {\bibinfo {volume} {33}},\ \bibinfo {pages} {2104083} (\bibinfo {year} {2021})}\BibitemShut {NoStop}%
\bibitem [{\citenamefont {Zeng}\ \emph {et~al.}(2022)\citenamefont {Zeng}, \citenamefont {Li}, \citenamefont {Chow}, \citenamefont {Cao}, \citenamefont {Zhang}, \citenamefont {Tang}, \citenamefont {Yin}, \citenamefont {Lim}, \citenamefont {Hu}, \citenamefont {Yang},\ and\ \citenamefont {Ariando}}]{zeng2022superconductivity}%
  \BibitemOpen
  \bibfield  {author} {\bibinfo {author} {\bibfnamefont {S.}~\bibnamefont {Zeng}}, \bibinfo {author} {\bibfnamefont {C.}~\bibnamefont {Li}}, \bibinfo {author} {\bibfnamefont {L.~E.}\ \bibnamefont {Chow}}, \bibinfo {author} {\bibfnamefont {Y.}~\bibnamefont {Cao}}, \bibinfo {author} {\bibfnamefont {Z.}~\bibnamefont {Zhang}}, \bibinfo {author} {\bibfnamefont {C.~S.}\ \bibnamefont {Tang}}, \bibinfo {author} {\bibfnamefont {X.}~\bibnamefont {Yin}}, \bibinfo {author} {\bibfnamefont {Z.~S.}\ \bibnamefont {Lim}}, \bibinfo {author} {\bibfnamefont {J.}~\bibnamefont {Hu}}, \bibinfo {author} {\bibfnamefont {P.}~\bibnamefont {Yang}},\ and\ \bibinfo {author} {\bibfnamefont {A.}~\bibnamefont {Ariando}},\ }\bibfield  {title} {\bibinfo {title} {{Superconductivity in infinite-layer nickelate La$_{1-x}$Ca${_x}$NiO$_2$ thin films}},\ }\href {https://doi.org/10.1126/sciadv.abl9927} {\bibfield  {journal} {\bibinfo  {journal} {Science Advances}\ }\textbf {\bibinfo {volume} {8}},\ \bibinfo {pages} {eabl9927} (\bibinfo {year}
  {2022})}\BibitemShut {NoStop}%
\bibitem [{\citenamefont {Wei}\ \emph {et~al.}(2023)\citenamefont {Wei}, \citenamefont {Vu}, \citenamefont {Zhang}, \citenamefont {Walker},\ and\ \citenamefont {Ahn}}]{Wei2023_NENO}%
  \BibitemOpen
  \bibfield  {author} {\bibinfo {author} {\bibfnamefont {W.}~\bibnamefont {Wei}}, \bibinfo {author} {\bibfnamefont {D.}~\bibnamefont {Vu}}, \bibinfo {author} {\bibfnamefont {Z.}~\bibnamefont {Zhang}}, \bibinfo {author} {\bibfnamefont {F.~J.}\ \bibnamefont {Walker}},\ and\ \bibinfo {author} {\bibfnamefont {C.~H.}\ \bibnamefont {Ahn}},\ }\bibfield  {title} {\bibinfo {title} {Superconducting {N}d$_{1-x}${E}u$_x${N}i{O}$_2$ thin films using in situ synthesis},\ }\href {https://doi.org/https://doi.org/10.1126/sciadv.adh3327} {\bibfield  {journal} {\bibinfo  {journal} {Science Advances}\ }\textbf {\bibinfo {volume} {9}},\ \bibinfo {pages} {eadh3327} (\bibinfo {year} {2023})}\BibitemShut {NoStop}%
\bibitem [{\citenamefont {Pan}\ \emph {et~al.}(2022)\citenamefont {Pan}, \citenamefont {Ferenc~Segedin}, \citenamefont {LaBollita}, \citenamefont {Song}, \citenamefont {Nica}, \citenamefont {Goodge}, \citenamefont {Pierce}, \citenamefont {Doyle}, \citenamefont {Novakov}, \citenamefont {C{\'o}rdova~Carrizales}, \citenamefont {N'Diaye}, \citenamefont {Shafer}, \citenamefont {Paik}, \citenamefont {Heron}, \citenamefont {Mason}, \citenamefont {Yacoby}, \citenamefont {Kourkoutis}, \citenamefont {Erten}, \citenamefont {Brooks}, \citenamefont {Botana},\ and\ \citenamefont {Mundy}}]{pan2022superconductivity}%
  \BibitemOpen
  \bibfield  {author} {\bibinfo {author} {\bibfnamefont {G.~A.}\ \bibnamefont {Pan}}, \bibinfo {author} {\bibfnamefont {D.}~\bibnamefont {Ferenc~Segedin}}, \bibinfo {author} {\bibfnamefont {H.}~\bibnamefont {LaBollita}}, \bibinfo {author} {\bibfnamefont {Q.}~\bibnamefont {Song}}, \bibinfo {author} {\bibfnamefont {E.~M.}\ \bibnamefont {Nica}}, \bibinfo {author} {\bibfnamefont {B.~H.}\ \bibnamefont {Goodge}}, \bibinfo {author} {\bibfnamefont {A.~T.}\ \bibnamefont {Pierce}}, \bibinfo {author} {\bibfnamefont {S.}~\bibnamefont {Doyle}}, \bibinfo {author} {\bibfnamefont {S.}~\bibnamefont {Novakov}}, \bibinfo {author} {\bibfnamefont {D.}~\bibnamefont {C{\'o}rdova~Carrizales}}, \bibinfo {author} {\bibfnamefont {A.~T.}\ \bibnamefont {N'Diaye}}, \bibinfo {author} {\bibfnamefont {P.}~\bibnamefont {Shafer}}, \bibinfo {author} {\bibfnamefont {H.}~\bibnamefont {Paik}}, \bibinfo {author} {\bibfnamefont {J.~T.}\ \bibnamefont {Heron}}, \bibinfo {author} {\bibfnamefont {J.~A.}\ \bibnamefont {Mason}}, \bibinfo {author}
  {\bibfnamefont {A.}~\bibnamefont {Yacoby}}, \bibinfo {author} {\bibfnamefont {L.~F.}\ \bibnamefont {Kourkoutis}}, \bibinfo {author} {\bibfnamefont {O.}~\bibnamefont {Erten}}, \bibinfo {author} {\bibfnamefont {C.~M.}\ \bibnamefont {Brooks}}, \bibinfo {author} {\bibfnamefont {A.~S.}\ \bibnamefont {Botana}},\ and\ \bibinfo {author} {\bibfnamefont {J.~A.}\ \bibnamefont {Mundy}},\ }\bibfield  {title} {\bibinfo {title} {Superconductivity in a quintuple-layer square-planar nickelate},\ }\href {https://doi.org/10.1038/s41563-021-01142-9} {\bibfield  {journal} {\bibinfo  {journal} {Nat. Mater.}\ }\textbf {\bibinfo {volume} {21}},\ \bibinfo {pages} {160} (\bibinfo {year} {2022})}\BibitemShut {NoStop}%
\bibitem [{\citenamefont {Kluge}\ \emph {et~al.}(1995)\citenamefont {Kluge}, \citenamefont {Koike}, \citenamefont {Fujiwara}, \citenamefont {Kato}, \citenamefont {Noji},\ and\ \citenamefont {Saito}}]{kluge1995clear}%
  \BibitemOpen
  \bibfield  {author} {\bibinfo {author} {\bibfnamefont {T.}~\bibnamefont {Kluge}}, \bibinfo {author} {\bibfnamefont {Y.}~\bibnamefont {Koike}}, \bibinfo {author} {\bibfnamefont {A.}~\bibnamefont {Fujiwara}}, \bibinfo {author} {\bibfnamefont {M.}~\bibnamefont {Kato}}, \bibinfo {author} {\bibfnamefont {T.}~\bibnamefont {Noji}},\ and\ \bibinfo {author} {\bibfnamefont {Y.}~\bibnamefont {Saito}},\ }\bibfield  {title} {\bibinfo {title} {Clear distinction between the underdoped and overdoped regime in the ${{\mathit{T}}_{\mathit{c}}}$ suppression of {Cu}-site-substituted high-${{\mathit{T}}_{\mathit{c}}}$ cuprates},\ }\href {https://doi.org/10.1103/PhysRevB.52.R727} {\bibfield  {journal} {\bibinfo  {journal} {Phys. Rev. B}\ }\textbf {\bibinfo {volume} {52}},\ \bibinfo {pages} {R727} (\bibinfo {year} {1995})}\BibitemShut {NoStop}%
\bibitem [{\citenamefont {Li}\ \emph {et~al.}(2020)\citenamefont {Li}, \citenamefont {Wang}, \citenamefont {Lee}, \citenamefont {Harvey}, \citenamefont {Osada}, \citenamefont {Goodge}, \citenamefont {Kourkoutis},\ and\ \citenamefont {Hwang}}]{Li2020_Dome}%
  \BibitemOpen
  \bibfield  {author} {\bibinfo {author} {\bibfnamefont {D.}~\bibnamefont {Li}}, \bibinfo {author} {\bibfnamefont {B.~Y.}\ \bibnamefont {Wang}}, \bibinfo {author} {\bibfnamefont {K.}~\bibnamefont {Lee}}, \bibinfo {author} {\bibfnamefont {S.~P.}\ \bibnamefont {Harvey}}, \bibinfo {author} {\bibfnamefont {M.}~\bibnamefont {Osada}}, \bibinfo {author} {\bibfnamefont {B.~H.}\ \bibnamefont {Goodge}}, \bibinfo {author} {\bibfnamefont {L.~F.}\ \bibnamefont {Kourkoutis}},\ and\ \bibinfo {author} {\bibfnamefont {H.~Y.}\ \bibnamefont {Hwang}},\ }\bibfield  {title} {\bibinfo {title} {{Superconducting Dome in Nd$_{0.825}$Sr$_{0.175}$NiO$_{2}$ Infinite Layer Films}},\ }\href {https://doi.org/10.1103/PhysRevLett.125.027001} {\bibfield  {journal} {\bibinfo  {journal} {Phys. Rev. Lett.}\ }\textbf {\bibinfo {volume} {125}},\ \bibinfo {pages} {027001} (\bibinfo {year} {2020})}\BibitemShut {NoStop}%
\bibitem [{\citenamefont {Zeng}\ \emph {et~al.}(2020)\citenamefont {Zeng}, \citenamefont {Tang}, \citenamefont {Yin}, \citenamefont {Li}, \citenamefont {Li}, \citenamefont {Huang}, \citenamefont {Hu}, \citenamefont {Liu}, \citenamefont {Omar}, \citenamefont {Jani}, \citenamefont {Lim}, \citenamefont {Han}, \citenamefont {Wan}, \citenamefont {Yang}, \citenamefont {Pennycook}, \citenamefont {Wee},\ and\ \citenamefont {Ariando}}]{zeng2020phase}%
  \BibitemOpen
  \bibfield  {author} {\bibinfo {author} {\bibfnamefont {S.}~\bibnamefont {Zeng}}, \bibinfo {author} {\bibfnamefont {C.~S.}\ \bibnamefont {Tang}}, \bibinfo {author} {\bibfnamefont {X.}~\bibnamefont {Yin}}, \bibinfo {author} {\bibfnamefont {C.}~\bibnamefont {Li}}, \bibinfo {author} {\bibfnamefont {M.}~\bibnamefont {Li}}, \bibinfo {author} {\bibfnamefont {Z.}~\bibnamefont {Huang}}, \bibinfo {author} {\bibfnamefont {J.}~\bibnamefont {Hu}}, \bibinfo {author} {\bibfnamefont {W.}~\bibnamefont {Liu}}, \bibinfo {author} {\bibfnamefont {G.~J.}\ \bibnamefont {Omar}}, \bibinfo {author} {\bibfnamefont {H.}~\bibnamefont {Jani}}, \bibinfo {author} {\bibfnamefont {Z.~S.}\ \bibnamefont {Lim}}, \bibinfo {author} {\bibfnamefont {K.}~\bibnamefont {Han}}, \bibinfo {author} {\bibfnamefont {D.}~\bibnamefont {Wan}}, \bibinfo {author} {\bibfnamefont {P.}~\bibnamefont {Yang}}, \bibinfo {author} {\bibfnamefont {S.~J.}\ \bibnamefont {Pennycook}}, \bibinfo {author} {\bibfnamefont {A.~T.~S.}\ \bibnamefont {Wee}},\ and\ \bibinfo {author}
  {\bibfnamefont {A.}~\bibnamefont {Ariando}},\ }\bibfield  {title} {\bibinfo {title} {{Phase Diagram and Superconducting Dome of Infinite-Layer ${{\mathrm{Nd}}_{1\ensuremath{-}x}{\mathrm{Sr}}_{x}{\mathrm{NiO}}_{2}}$ Thin Films}},\ }\href {https://doi.org/10.1103/PhysRevLett.125.147003} {\bibfield  {journal} {\bibinfo  {journal} {Phys. Rev. Lett.}\ }\textbf {\bibinfo {volume} {125}},\ \bibinfo {pages} {147003} (\bibinfo {year} {2020})}\BibitemShut {NoStop}%
\bibitem [{\citenamefont {Cervasio}\ \emph {et~al.}(2023)\citenamefont {Cervasio}, \citenamefont {Tomarchio}, \citenamefont {Verseils}, \citenamefont {Brubach}, \citenamefont {Macis}, \citenamefont {Zeng}, \citenamefont {Ariando}, \citenamefont {Roy},\ and\ \citenamefont {Lupi}}]{cervasio2023optical}%
  \BibitemOpen
  \bibfield  {author} {\bibinfo {author} {\bibfnamefont {R.}~\bibnamefont {Cervasio}}, \bibinfo {author} {\bibfnamefont {L.}~\bibnamefont {Tomarchio}}, \bibinfo {author} {\bibfnamefont {M.}~\bibnamefont {Verseils}}, \bibinfo {author} {\bibfnamefont {J.-B.}\ \bibnamefont {Brubach}}, \bibinfo {author} {\bibfnamefont {S.}~\bibnamefont {Macis}}, \bibinfo {author} {\bibfnamefont {S.}~\bibnamefont {Zeng}}, \bibinfo {author} {\bibfnamefont {A.}~\bibnamefont {Ariando}}, \bibinfo {author} {\bibfnamefont {P.}~\bibnamefont {Roy}},\ and\ \bibinfo {author} {\bibfnamefont {S.}~\bibnamefont {Lupi}},\ }\bibfield  {title} {\bibinfo {title} {{Optical Properties of Superconducting {Nd$_{0.8}$Sr$_{0.2}$NiO$_2$} Nickelate}},\ }\href {https://doi.org/10.1021/acsaelm.3c00506} {\bibfield  {journal} {\bibinfo  {journal} {ACS Applied Electronic Materials}\ }\textbf {\bibinfo {volume} {5}},\ \bibinfo {pages} {4770} (\bibinfo {year} {2023})}\BibitemShut {NoStop}%
\bibitem [{\citenamefont {Ding}\ \emph {et~al.}(2024)\citenamefont {Ding}, \citenamefont {Fan}, \citenamefont {Wang}, \citenamefont {Li}, \citenamefont {An}, \citenamefont {Ye}, \citenamefont {Tang}, \citenamefont {Lei}, \citenamefont {Sun}, \citenamefont {Guo}, \citenamefont {Chen}, \citenamefont {Sangphet}, \citenamefont {Wang}, \citenamefont {Xu}, \citenamefont {Peng},\ and\ \citenamefont {Feng}}]{ding2024cuprate}%
  \BibitemOpen
  \bibfield  {author} {\bibinfo {author} {\bibfnamefont {X.}~\bibnamefont {Ding}}, \bibinfo {author} {\bibfnamefont {Y.}~\bibnamefont {Fan}}, \bibinfo {author} {\bibfnamefont {X.}~\bibnamefont {Wang}}, \bibinfo {author} {\bibfnamefont {C.}~\bibnamefont {Li}}, \bibinfo {author} {\bibfnamefont {Z.}~\bibnamefont {An}}, \bibinfo {author} {\bibfnamefont {J.}~\bibnamefont {Ye}}, \bibinfo {author} {\bibfnamefont {S.}~\bibnamefont {Tang}}, \bibinfo {author} {\bibfnamefont {M.}~\bibnamefont {Lei}}, \bibinfo {author} {\bibfnamefont {X.}~\bibnamefont {Sun}}, \bibinfo {author} {\bibfnamefont {N.}~\bibnamefont {Guo}}, \bibinfo {author} {\bibfnamefont {Z.}~\bibnamefont {Chen}}, \bibinfo {author} {\bibfnamefont {S.}~\bibnamefont {Sangphet}}, \bibinfo {author} {\bibfnamefont {Y.}~\bibnamefont {Wang}}, \bibinfo {author} {\bibfnamefont {H.}~\bibnamefont {Xu}}, \bibinfo {author} {\bibfnamefont {R.}~\bibnamefont {Peng}},\ and\ \bibinfo {author} {\bibfnamefont {D.}~\bibnamefont {Feng}},\ }\bibfield  {title} {\bibinfo {title}
  {{Cuprate-like electronic structures in infinite-layer nickelates with substantial hole dopings}},\ }\href {https://doi.org/10.1093/nsr/nwae194} {\bibfield  {journal} {\bibinfo  {journal} {National Science Review}\ }\textbf {\bibinfo {volume} {11}},\ \bibinfo {pages} {nwae194} (\bibinfo {year} {2024})}\BibitemShut {NoStop}%
\bibitem [{\citenamefont {Sun}\ \emph {et~al.}(2025)\citenamefont {Sun}, \citenamefont {Jiang}, \citenamefont {Xia}, \citenamefont {Hao}, \citenamefont {Yan}, \citenamefont {Wang}, \citenamefont {Li}, \citenamefont {Liu}, \citenamefont {Ding}, \citenamefont {Liu}, \citenamefont {Liu}, \citenamefont {Liu}, \citenamefont {Chen}, \citenamefont {Shen},\ and\ \citenamefont {Nie}}]{Sun2025_LSNO_ARPES}%
  \BibitemOpen
  \bibfield  {author} {\bibinfo {author} {\bibfnamefont {W.}~\bibnamefont {Sun}}, \bibinfo {author} {\bibfnamefont {Z.}~\bibnamefont {Jiang}}, \bibinfo {author} {\bibfnamefont {C.}~\bibnamefont {Xia}}, \bibinfo {author} {\bibfnamefont {B.}~\bibnamefont {Hao}}, \bibinfo {author} {\bibfnamefont {S.}~\bibnamefont {Yan}}, \bibinfo {author} {\bibfnamefont {M.}~\bibnamefont {Wang}}, \bibinfo {author} {\bibfnamefont {Y.}~\bibnamefont {Li}}, \bibinfo {author} {\bibfnamefont {H.}~\bibnamefont {Liu}}, \bibinfo {author} {\bibfnamefont {J.}~\bibnamefont {Ding}}, \bibinfo {author} {\bibfnamefont {J.}~\bibnamefont {Liu}}, \bibinfo {author} {\bibfnamefont {Z.}~\bibnamefont {Liu}}, \bibinfo {author} {\bibfnamefont {J.}~\bibnamefont {Liu}}, \bibinfo {author} {\bibfnamefont {H.}~\bibnamefont {Chen}}, \bibinfo {author} {\bibfnamefont {D.}~\bibnamefont {Shen}},\ and\ \bibinfo {author} {\bibfnamefont {Y.}~\bibnamefont {Nie}},\ }\bibfield  {title} {\bibinfo {title} {Electronic structure of superconducting infinite-layer lanthanum
  nickelates},\ }\href {https://doi.org/10.1126/sciadv.adr5116} {\bibfield  {journal} {\bibinfo  {journal} {Science Advances}\ }\textbf {\bibinfo {volume} {11}},\ \bibinfo {pages} {eadr5116} (\bibinfo {year} {2025})}\BibitemShut {NoStop}%
\bibitem [{\citenamefont {Wang}\ \emph {et~al.}(2021)\citenamefont {Wang}, \citenamefont {Li}, \citenamefont {Goodge}, \citenamefont {Lee}, \citenamefont {Osada}, \citenamefont {Harvey}, \citenamefont {Kourkoutis}, \citenamefont {Beasley},\ and\ \citenamefont {Hwang}}]{wang2021isotropic}%
  \BibitemOpen
  \bibfield  {author} {\bibinfo {author} {\bibfnamefont {B.~Y.}\ \bibnamefont {Wang}}, \bibinfo {author} {\bibfnamefont {D.}~\bibnamefont {Li}}, \bibinfo {author} {\bibfnamefont {B.~H.}\ \bibnamefont {Goodge}}, \bibinfo {author} {\bibfnamefont {K.}~\bibnamefont {Lee}}, \bibinfo {author} {\bibfnamefont {M.}~\bibnamefont {Osada}}, \bibinfo {author} {\bibfnamefont {S.~P.}\ \bibnamefont {Harvey}}, \bibinfo {author} {\bibfnamefont {L.~F.}\ \bibnamefont {Kourkoutis}}, \bibinfo {author} {\bibfnamefont {M.~R.}\ \bibnamefont {Beasley}},\ and\ \bibinfo {author} {\bibfnamefont {H.~Y.}\ \bibnamefont {Hwang}},\ }\bibfield  {title} {\bibinfo {title} {{Isotropic Pauli-limited superconductivity in the infinite-layer nickelate {Nd}$_{0.775}${Sr}$_{0.225}${NiO}$_2$}},\ }\href {https://doi.org/10.1038/s41567-020-01128-5} {\bibfield  {journal} {\bibinfo  {journal} {Nature Physics}\ }\textbf {\bibinfo {volume} {17}},\ \bibinfo {pages} {473} (\bibinfo {year} {2021})}\BibitemShut {NoStop}%
\bibitem [{\citenamefont {Boschini}\ \emph {et~al.}(2013)\citenamefont {Boschini}, \citenamefont {Consolandi}, \citenamefont {Gervasi}, \citenamefont {Giani}, \citenamefont {Grandi}, \citenamefont {Ivanchenko}, \citenamefont {Nieminem}, \citenamefont {Pensotti}, \citenamefont {Rancoita},\ and\ \citenamefont {Tacconi}}]{boschini2013expression}%
  \BibitemOpen
  \bibfield  {author} {\bibinfo {author} {\bibfnamefont {M.}~\bibnamefont {Boschini}}, \bibinfo {author} {\bibfnamefont {C.}~\bibnamefont {Consolandi}}, \bibinfo {author} {\bibfnamefont {M.}~\bibnamefont {Gervasi}}, \bibinfo {author} {\bibfnamefont {S.}~\bibnamefont {Giani}}, \bibinfo {author} {\bibfnamefont {D.}~\bibnamefont {Grandi}}, \bibinfo {author} {\bibfnamefont {V.}~\bibnamefont {Ivanchenko}}, \bibinfo {author} {\bibfnamefont {P.}~\bibnamefont {Nieminem}}, \bibinfo {author} {\bibfnamefont {S.}~\bibnamefont {Pensotti}}, \bibinfo {author} {\bibfnamefont {P.}~\bibnamefont {Rancoita}},\ and\ \bibinfo {author} {\bibfnamefont {M.}~\bibnamefont {Tacconi}},\ }\bibfield  {title} {\bibinfo {title} {{An expression for the Mott cross section of electrons and positrons on nuclei with Z up to 118}},\ }\href {https://www.sciencedirect.com/science/article/pii/S0969806X13002454} {\bibfield  {journal} {\bibinfo  {journal} {Radiation Physics and Chemistry}\ }\textbf {\bibinfo {volume} {90}},\ \bibinfo {pages} {39}
  (\bibinfo {year} {2013})}\BibitemShut {NoStop}%
\bibitem [{\citenamefont {McGuinness}(2022)}]{mcguinness2022high}%
  \BibitemOpen
  \bibfield  {author} {\bibinfo {author} {\bibfnamefont {P.~H.}\ \bibnamefont {McGuinness}},\ }\bibinfo {title} {High energy electron irradiation of delafossite metals},\ in\ \href {https://doi.org/10.1007/978-3-031-14244-4_4} {\emph {\bibinfo {booktitle} {Probing Unconventional Transport Regimes in Delafossite Metals}}}\ (\bibinfo  {publisher} {Springer International Publishing},\ \bibinfo {year} {2022})\ pp.\ \bibinfo {pages} {41--86}\BibitemShut {NoStop}%
\end{thebibliography}%

\newpage
\onecolumngrid

\renewcommand{\thefigure}{S\arabic{figure}}

\section{Supplemental Information}

\subsection{Irradiation-induced defects}
% *some comment on this, can also include the cross section calculations here*

High energy electron irradiation adds disorder in a well controlled manner. 
Usually electrons in the megavolt energy range are used for such cases. 
At such energies, electrons travel at relativistic speeds and due to their small relative mass, energy transfer to the much heavier nuclei in the target material remains quite low. 
This causes the creation of isolated point defects, generally considered as a vacancy-interstitial pair, which can create localized energy levels in the electronic band structure and also act as scattering centers \cite{Alessi2023_Electron_irradiation}. 
Such high energy electrons have a large penetration depth (on the order of mm) which ensures a homogeneous irradiation throughout the entire depths of the samples used here (NSNO film thicknesses $\sim 5$ nm and total sample thicknesses $<300$ {\textmu}m). 

\begin{figure}[H]
    \centering
    \includegraphics[width = \columnwidth]{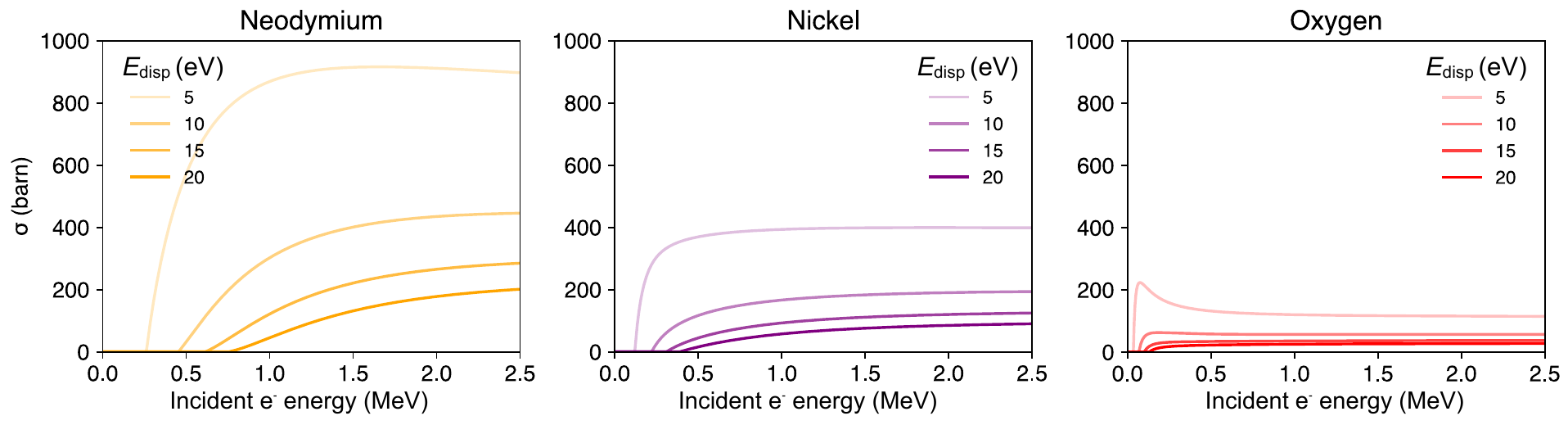}
\caption{Calculated scattering cross sections of the constituent atoms for a range of removal energies. }
    \label{SIfig_scattering}
\end{figure}

The relative probability of creating a defect at any atomic site can be predicted by calculating the scattering cross section for a given element, primary (incident) electron energy, and the energy $E_{disp}$ required to displace the ion from its native lattice position \cite{boschini2013expression, mcguinness2022high}. 
Calculated cross sections for the atomic species studied here -- Nd, Ni, and O -- as a function of primary voltage are shown in Figure \ref{SIfig_scattering} for a range of displacement energies. 
% \emph{Ab initio} calculations by density functional theory (DFT) and the nudged elastic band (NEB) model \cite{} indicate that the displacement energy for a Ni atom is $\sim$15 eV. 
In general, collisions with heavier ions are expected to dominate at high irradiation energies, as is visible from the relative cross sections of Nd and Ni compared to O for the same removal energies. 
Typical removal energies for similar materials are often estimated in the range of $\sim$10-25 eV \cite{sunko2020controlled, Alessi2023_Electron_irradiation}.

% \begin{figure}[H]
%     \centering
%     \includegraphics[width = 0.7\columnwidth]{SI_figures/SIfig_NEB.pdf}
% \caption{Energy barrier for the formation of a (0,1,1/2) Ni interstitial site. The evolution of the system between the initial (native) and final (Ni interstitial) structure are shown in the upper panels in perspective and from a $c$-axis projection.}
%     \label{SIfig_NEB}
% \end{figure}

\textcolor{darkgreen}{Generally, the resulting point-like defects are often considered non-magnetic. 
The irradiation process displaces atoms without introducing foreign magnetic impurities, such that the vacancy-interstitial pairs created typically do not give rise to additional localized magnetic moments, even in cases when the constituent ions could be magnetic in other contexts.
For example, magnetic measurements following similar irradiation experiments of superconducting materials constituted of classically magnetic atoms, including
\ch{LaNiGa2} \cite{Ghimire2024_LaNiGa2}, iron-based superconductors such as $\mathrm{Ba(Fe_{1-x}Ru_x)_2As_2}$ \cite{prozorov2014effect},  $\mathrm{BaFe_2(As_{1-x}P_x)_2}$\cite{Mizukami2014_iron_pnictide}, and  $\mathrm{SrTi_{1-x}Nb_xO_3}$\cite{Xiao2015_dopedSTO} have indicated the non-magnetic nature of these defects. 
}
\textcolor{darkgreen}{
While experimental measurements of magnetic ions in thin films with small sample volume fractions present numerous challenges for robust and quantitative interpretation, preliminary investigations by two-coil mutual inductance before and after electron irradiation showed no signs of additional magnetic contributions in the films studied here.  
}

% {\color{darkgreen}{
% The standard experimental techniques to measure magnetic properties of materials are very challenging to perform on thin films due to small sample volumes compared to the extremely large volume of substrate material contributing to the background. \cite{Ney2008_limitations_magnetism}. 

% \textbf{Calculations? }

% Expected total magnetic moment arising due to irradiation (the extremely small number of defects produced $<1\%$) lies at the borderline of the most sensitive commercially available SQUID based magnetometers. This coupled with standard methods of large diamagnetic background subtraction (substrate signal) makes robust and quantitative interpretation to probe post irradiation magnetism (even if present at all) extremely hard \cite{Buchner2018_SQUID_tutorial}.
% % Experimental measurements of magnetic ions in thin films (small sample volume fractions) such as those studied here present numerous challenges for robust and quantitative interpretation \cite{}. 
% % - why is SQUID hard, with references 
% % - penetration depth, with references
% % - susceptibility, with references
% % - why is ESR? hard, with references
% Here, preliminary investigations by electron spin resonance (ESR) and two-coil mutual inductance before and after electron irradiation showed no signs of additional magnetic contributions in the films studied.  
% }}

\pagebreak

\subsection{Resistivity measurements}
Temperature-dependent resistivity measurements were performed using the four-probe method, with aluminum wire-bonded contacts.
Voltage drop was measured either on a lock-in amplifier (SYNKTEK) with an excitation frequency of 117.77 Hz or a nanovoltmeter (Keithley 2182A), where the polarity of the current was periodically flipped to reduce the contributions from thermal EMF effects. 
Samples were biased with currents of $<1$ \textmu A in all cases.

\begin{figure}[H]
    \centering
    \includegraphics[width = 0.5\columnwidth]{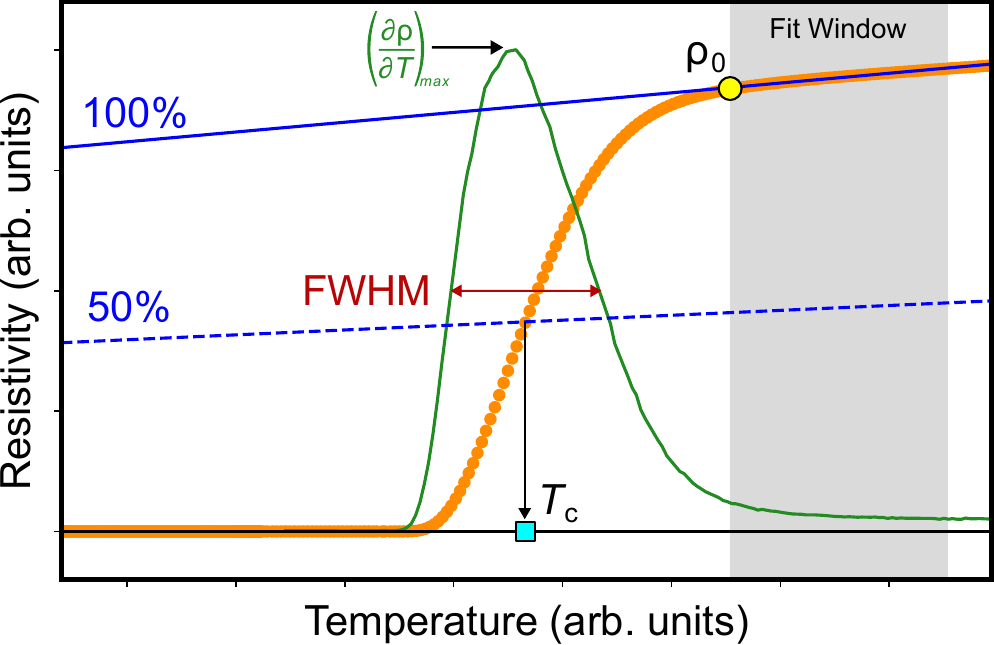}
\caption{Schematic for parameter extraction from $\rho(T)$ data. 
The first derivative (green) of the measured $\rho (T)$ (orange) is taken and the full-width at half-maximum (FWHM) is calculated. 
A 5 K fit window (grey shading) is defined starting at $1.5\times\mathrm{FWHM}$ from the peak of $\pdv{\rho}/{T}$. 
Residual resistivity $\rho_0$ is the value at the lowest temperature of the fit window.
The linear fit estimated from the window is considered 100\% of the superconducting transition. The temperature at which the 50\% of the fit-line (dashed blue line) intersects the $\rho(T)$ data is assigned to be the {\Tc} for the particular measurement.
The superconducting transition width $\delta T_c$ is the FWHM.}
    \label{SIfig_params}
\end{figure}

\pagebreak

\begin{figure}[H]
    \centering
    \includegraphics[width = \columnwidth]{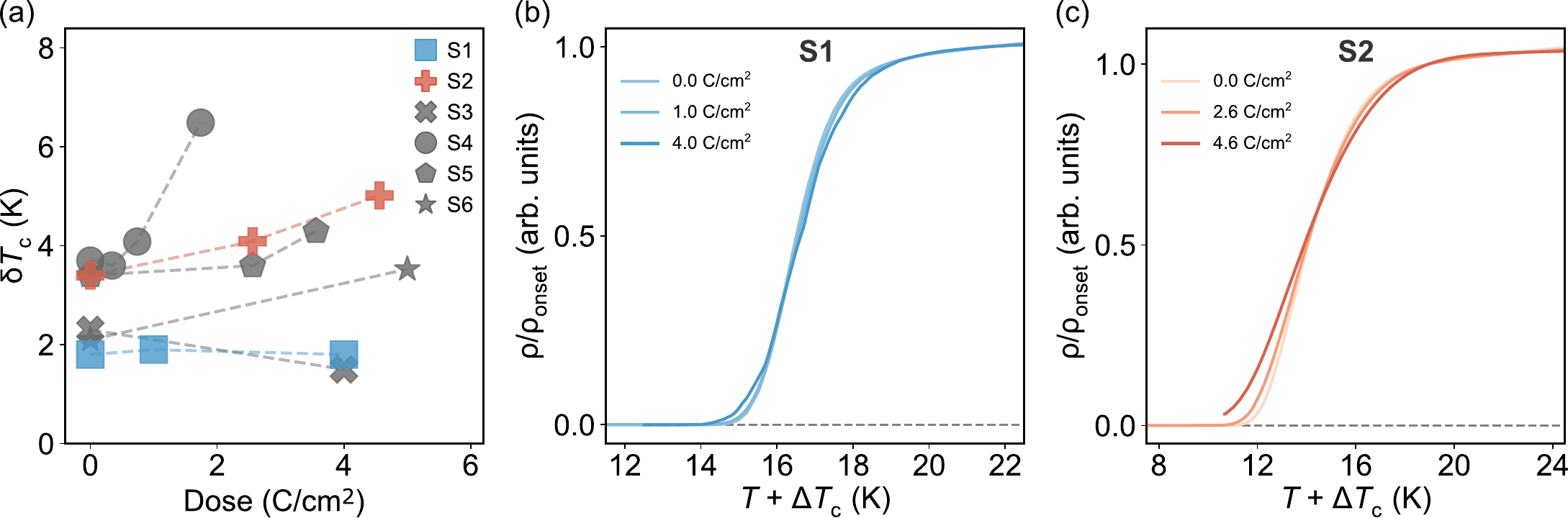}
\caption{(a) Evolution of superconducting transition width $\delta T_c$ with progressive irradiation dose. Samples S1 and S2 at their maximum doses (\dose{8.0} and \dose{6.6} respectively) do not show complete transitions such that the transition widths cannot be determined and the corresponding data points are excluded from this plot. Resistivity versus temperature plots normalized to $\rho_{\mathrm{onset}}$ and horizontally shifted by the corresponding $\Delta T_c$ for (a) S1 and (b) S2, demonstrating that the sharpness of the superconducting transitions remain essentially unchanged over varying levels of irradiation-induced disorder.}
    \label{SIfig_dose_width}
\end{figure} 

\begin{figure}[H]
    \centering
    \includegraphics[width = \columnwidth]{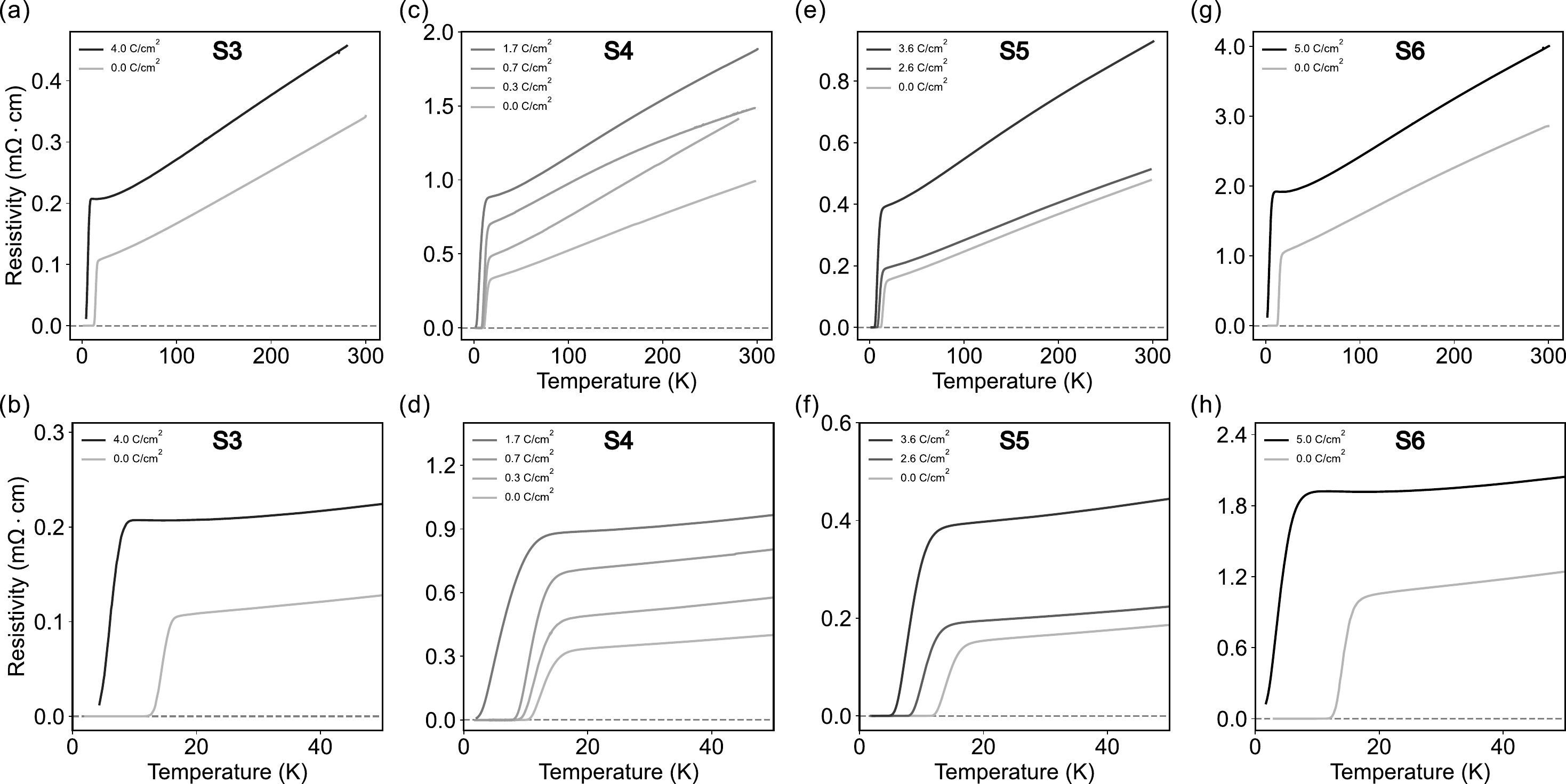}
\caption{Temperature-dependent resistivity measurements for additional samples S3-S6 with different irradiation doses. Sample S3 is cut from the same precursor film and reduction as sample S1 discussed in the main text; sample S4 equivalently for sample S2. Low-temperature insets are shown for each sample in the bottom row. }
    \label{SIfig_allRvT}
\end{figure}

\begin{figure}[H]
    \centering
    \includegraphics[width = \columnwidth]{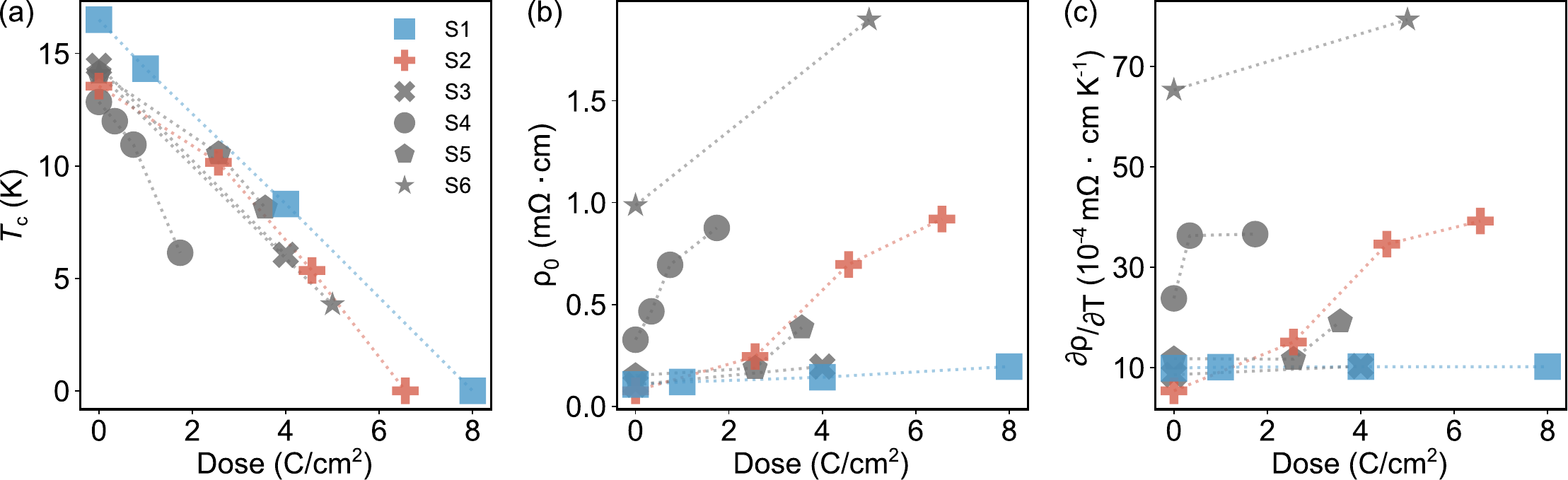}
\caption{Effect of disorder on (a) superconducting transition temperature, {\Tc}, (b) residual resistivity, $\rho_0$, and (c) normal state $T$-linear slope, $\pdv{\rho}/{T}$. There is notable sample-to-sample variation in the disorder evolution of both the normal state parameters. Very generally, samples showing a higher rate of $\rho_0$ increase also exhibit higher rates of normal state slope increase with added disorder.}
    \label{SIfig_rho0-slope}
\end{figure}

\subsection{Electronic transport under magnetic field}
\begin{figure}[H]
    \centering
    \includegraphics[width = 0.40\columnwidth]{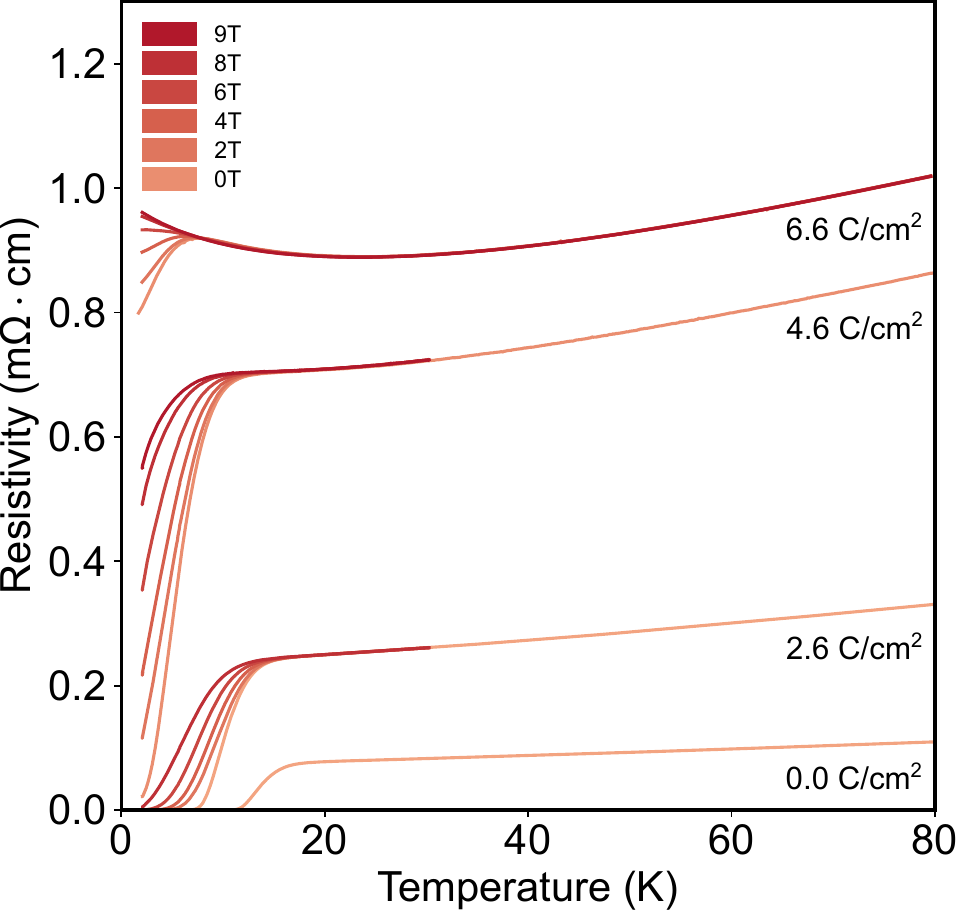}
\caption{Resistivity measurements of S2 in perpendicular (out of plane) magnetic fields. All measurements show a standard field suppression of {\Tc}. A resistive upturn is observed for the measurements corresponding to \dose{4.6 and 6.6}. The upturn temperature is not affected by the applied field.}
    \label{SIfig_field}
\end{figure}

\subsection{STEM Imaging}
\begin{figure}[H]
    \centering
    \includegraphics[width = 0.8\columnwidth]{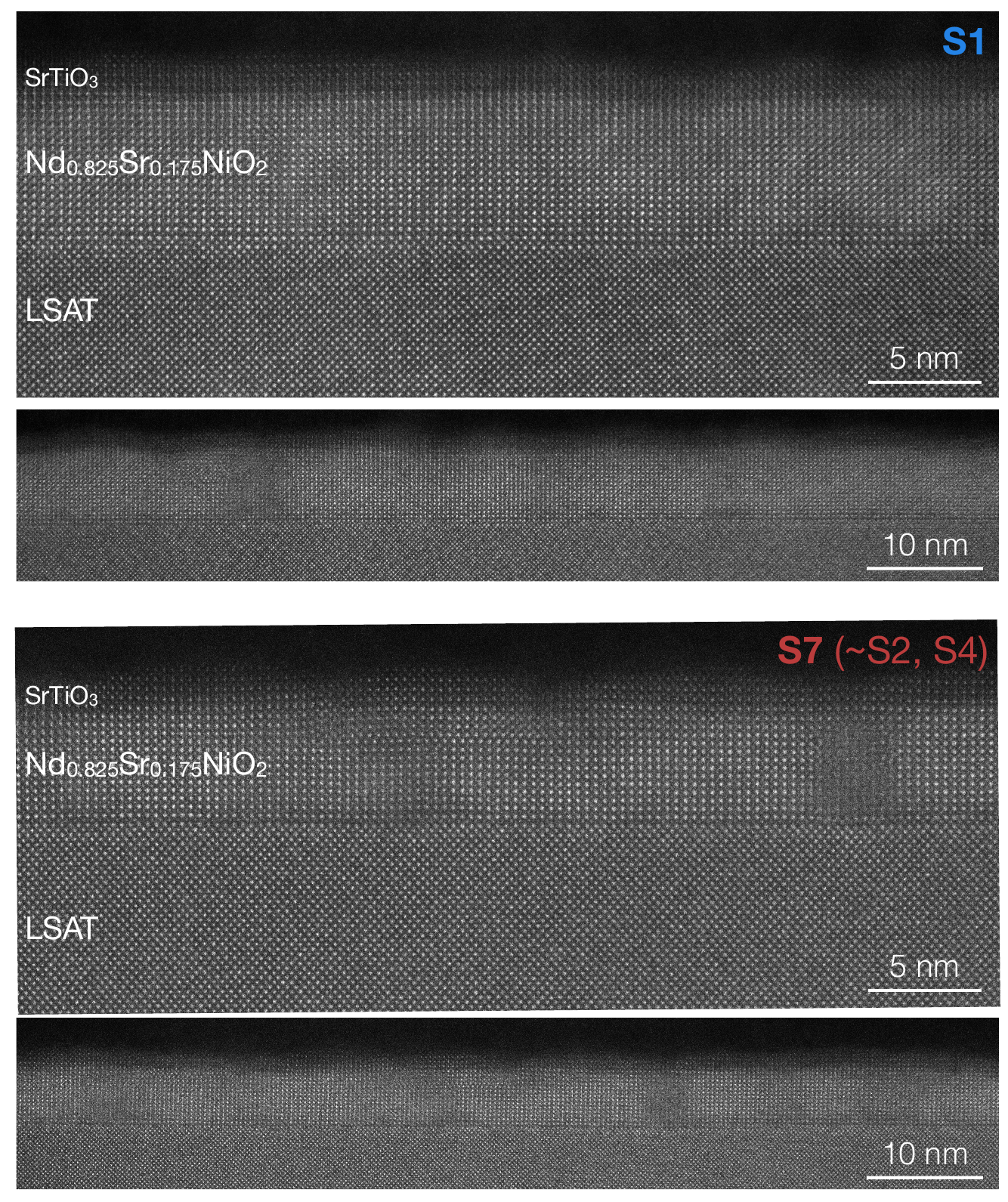}
\caption{Atomic structure of superconducting nickelate films used in this study investigated by high-angle annular dark-field scanning transmission electron microscopy (HAADF-STEM) imaging. Top: sample S1 discussed in the main text prior to high energy electron irradiation (dose = 0 C/cm$^2$); bottom: sample S7 cut from the same precursor growth and topotactic reduction as samples S2 and S4 after minimal irradiation (dose = 0.6 C/cm$^2$). Cross-sectional specimens were prepared by standard focussed ion beam lift-out and imaged in a double aberration-corrected JEOL ARM 300 operating at 300 kV with a probe convergence angle of 21 mrad. }
    \label{SIfig_STEM}
\end{figure}

{\color{darkgreen}
\subsection{X Ray Diffraction}

\begin{figure}[H]
    \centering
    \includegraphics[width=0.5\columnwidth]{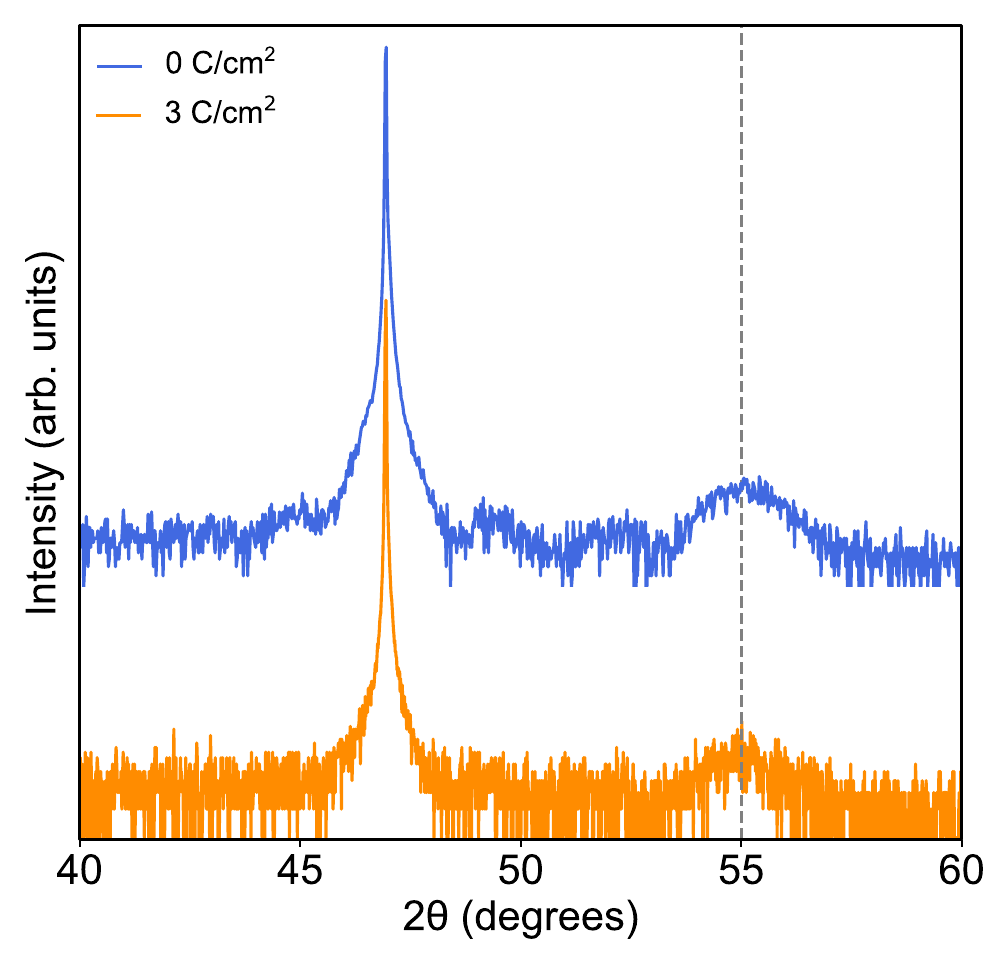}
    {\color{darkgreen}
    \caption{X-ray diffraction ($\lambda$: Cu K$\alpha$) spectra of an optimally doped Nd$_{0.825}$Sr$_{0.175}$NiO$_2$ thin film in the pristine (0 C/cm$^2$) and irradiated (3 C/cm$^2$ with 2.5 MeV electrons) state. The unchanged peak position at 2$\theta$ $\sim$ 55$^\circ$ confirms the preservation of the infinite-layer structure and the absence of spurious phases following irradiation.}}
    \label{SIfig_XRD}
\end{figure}
}

% \bibliography{references}

\end{document}